\documentclass[10pt,letterpaper]{article}

\usepackage[top=0.85in,left=2.75in,footskip=0.75in,marginparwidth=2in]{geometry}
\usepackage{amsmath,amssymb,bm}

\usepackage[utf8]{inputenc}
\usepackage{textgreek}

\usepackage{graphicx}          
\usepackage{caption}           
\usepackage[justification=justified,aboveskip=1pt,labelfont=bf,labelsep=period,singlelinecheck=off]{caption}
\usepackage{sidecap}           
\usepackage{wrapfig}           

\usepackage{tabularx,array,booktabs,makecell}
\usepackage{changepage}        

\usepackage[backend=biber,style=authoryear,sorting=nyt]{biblatex} 
\usepackage{hyperref,nameref}

\usepackage[right]{lineno}

\usepackage[expansion=false]{microtype}
\DisableLigatures[f]{encoding = *, family = * }

\usepackage{ragged2e}  
\justifying              

\usepackage{fancyhdr,lastpage}
\fancyheadoffset[L]{2.25in}
\fancyfootoffset[L]{2.25in}

\usepackage{color}
\definecolor{Gray}{gray}{.25}

\usepackage[pscoord]{eso-pic}
\usepackage[fulladjust]{marginnote}
\reversemarginpar

\begin{document}

\vspace*{0.35in}

\begingroup
\centering

{\LARGE Interpretable Electrophysiological Features of Resting-State EEG Capture Cortical Network Dynamics in Parkinson’s Disease\par}
\vspace{2ex}  

Antonios G. Dougalis\textsuperscript{1*}


\textsuperscript{1}Independent Researcher, 50131, Lefkopigi, Greece 

\vspace{1ex}  
* Correspondence, antoniosdougalis@med.uoc.gr; antoniosdougalis@gmail.com, 
orcID: 0000-0002-2139-1616

\vspace{2ex}  

\justifying

\section*{Abstract}
Background/Objectives: Parkinson’s disease (PD) alters cortical neural dynamics, yet reliable non-invasive electrophysiological biomarkers remain elusive. Resting-state electroencephalography (EEG) can capture complementary spectral, connectivity, and dynamical properties reflecting disease-related neural alterations and medication effects. This study examined whether interpretable EEG features capturing complementary aspects of neural dynamics can discriminate Parkinsonian neural states and provide insight into underlying electrophysiological differences. Methods: Resting-state EEG recordings from healthy controls and PD patients recorded in both off- and on-medication states were analyzed. A comprehensive set of interpretable features was extracted from each electrode and grouped into Standard descriptors (spectral power, phase synchronization, time-domain statistics) and Dynamical descriptors (aperiodic activity, cross-frequency coupling, scale-free dynamics, neuronal avalanche statistics, and instantaneous frequency measures). A multi-head attention transformer classifier was trained using strict leave-one-subject-out validation. Feature complementarity was assessed using random ablation analyses and pairwise correlation measures. Group-level comparisons were performed to identify electrophysiological differences associated with disease and medication state. Results: Standard and Dynamical feature sets achieved comparable classification performance. Standard features showed strongest performance in discriminating medication states (PDoff vs PDon), whereas Dynamical features performed competitively in contrasts between PD patients and healthy controls. Random feature ablation analyses indicated that Dynamical descriptors provide complementary information distributed across features while correlation analysis revealed low redundancy within both feature sets. Group-level comparisons revealed medication-sensitive reductions in delta power and voltage variance, modulation of neuronal avalanche statistics, persistent increases in theta phase synchronization in PD patients, and disease-related alterations in cross-frequency interactions. Conclusions: Distributed electrophysiological descriptors derived from resting-state EEG capture complementary aspects of Parkinsonian neural dynamics. Traditional spectral and synchronization features primarily reflect medication-related neural modulation, whereas dynamical descriptors reveal broader alterations in cortical network organization associated with disease but also with medication. These findings support multivariate EEG representations as a promising framework for developing non-invasive biomarkers of Parkinson’s disease.

\section*{Keywords}
Neural dynamics; Transformer models; Functional connectivity; Cross-frequency coupling; Leave-one-subject-out validation; Neuronal avalanches

\section*{Introduction}
Parkinson's disease (PD) is the second most common neurodegenerative disorder and it is the fastest growing neurological disease in prevalence and disability worldwide (Kowal \textit{et al.,} 2013, GBD 2016). The most recent projection data from the modelling study of Global Burden of Disease Study 2021 (Su \textit{et al.,} 2025) estimate around 25 million PD patients living worldwide by 2050 (112\% increase from 2021) with a forecasted prevalence of 267 per 100,000 of population (a rise of 76\% from 2021). The increase is attributed primarily to population ageing (89\%) followed by population growth (20\%) and changes in disease prevalence (3\%) (Su \textit{et al.,} 2025).

PD is characterized by the accumulation of $\alpha$-synuclein aggregates and progressive motor impairment resulting from degeneration of dopaminergic neurons within the nigrostriatal pathway and destabilization of neuronal activity in the basal ganglia (Dickson, 2012, Stocchi \textit{et al.,} 2024). The disease exhibits substantial heterogeneity in clinical phenotype, disease progression, and treatment response. Early and accurate diagnosis remains challenging, particularly during the prodromal and early symptomatic stages, since clinical examination alone may lack sufficient sensitivity while PET (Positron Emission Tomography) and SPECT (Single-Photon Emission Computed Tomography) imaging are expensive, often unavailable and, more importantly, unsuitable for repeated longitudinal assessment due to radiation exposure (Brucke \textit{et al.,} 2020; Waninger \textit{et al.,} 2020; Stocchi \textit{et al.,} 2024). Thus, identification of reliable non-invasive biomarkers capable of detecting PD pathology and tracking disease progression would have substantial clinical value, particularly for evaluating the effectiveness of emerging disease-modifying therapies in clinical trials (Stocchi \textit{et al.,} 2024).

Electroencephalography (EEG), especially during resting-state scalp EEG recordings, is a promising noninvasive candidate modality because it can capture cortical oscillatory dynamics which can be directly influenced by basal ganglia–thalamo–cortical circuits. For instance, elevated phase-amplitude coupling (PAC) between beta (13-30Hz) and broadband gamma bands (50-150Hz) registered in invasive electrocorticography (ECoG) in the motor cortex (de Hemptinne \textit{et al.,} 2013) is responsive to deep brain stimulation treatment (de Hemptinne \textit{et al.,} 2015) and can be detected in scalp EEG recordings from electrodes nearest the motor cortex (Swann \textit{et al.,} 2015). Similarly, features of beta oscillatory band waveform shape recorded over the somatosensory cortex exhibited distinct sharpness asymmetry and responded to changes in PD medication in resting-state EEG studies in PD patients (Jackson \textit{et al.,} 2019) Furthermore, scalp EEG recordings have detected widespread cortical increase in beta (13-30 Hz) and gamma(30-45Hz) band coherence in PD patients against controls that were positively correlated with disease severity markers and impaired dopaminergic tone measured by dopamine transporter PET imaging (Waninger \textit{et al.,} 2020) 

Despite these promising observations, consistent EEG biomarkers for PD remain elusive. Many conventional EEG metrics fail to reliably differentiate patients from healthy individuals at the global scalp level or at a localized level near the somatosensory cortex. For instance, studies have reported no consistent differences on beta and gamma band power in scalp EEG recordings (de Hemptinne \textit{et al.,} 2013; George \textit{et al.,} 2013; de Hemptinne \textit{et al.,} 2015; Swann \textit{et al.,} 2015), although changes in theta and alpha bands in posterior lobes (parietal \& occipital) have been detected in PD patients (Babiloni \textit{et al.,} 2011). These findings highlight that detectable EEG signatures of PD may be subtle, spatially heterogeneous, clinically divergent and highly dependent on the particular signal feature examined. These observations also suggest that single-feature EEG analyses may capture only partial aspects of Parkinsonian neural dynamics. The reliance on narrowly defined spatial or spectral features poses a challenge for biomarker development. 

An alternative approach is to treat the EEG not as a source of single candidate biomarkers, but rather as a high-dimensional representation of neural dynamics from which multiple complementary features can be extracted. In the present study, a multivariate perspective is adopted to investigate EEG signatures in Parkinson’s disease. Using a publicly available resting-state EEG database from healthy controls and Parkinson’s disease patients recorded in both on- and off-medication states (George \textit{et al.,} 2013), a comprehensive set of physiologically interpretable electrophysiological features was extracted from each electrode. These features encompass oscillatory and aperiodic spectral properties, temporal signal statistics, functional connectivity metrics, scale-free dynamics, excitation–inhibition(E/I) balance proxies, cross-frequency coupling measures, neuronal avalanche characteristics, and metrics of criticality dynamics. Rather than focusing on isolated candidate biomarkers, this approach treats EEG recordings as high-dimensional representations of neural activity from which complementary physiological descriptors of brain dynamics can be derived.

To evaluate whether these distributed electrophysiological descriptors contain discriminative information about Parkinsonian neural states, an attention-head transformer encoder model was utilized, a deep learning architecture increasingly applied in EEG feature integration and classification (Lyu \textit{et al.,} 2026). In addition to assessing classification performance, group-level analyses of the extracted features were conducted to examine how specific electrophysiological descriptors differ between healthy controls and Parkinson’s disease patients and how they are modulated by dopaminergic medication. These statistical comparisons were performed independently of the model-training procedure and serve to provide physiological interpretation of the feature representations. The extracted features were organized into two conceptual groups: \textit{standard features}, including spectral power, time-domain statistical descriptors, and phase-based functional connectivity measures; and \textit{dynamical features}, capturing higher-order properties of neural signal organization such as scale-free dynamics, cross-frequency interactions, and neuronal avalanche statistics. Classification performance was evaluated across both binary and three-class tasks distinguishing healthy controls, PD patients off medication, and the same patients during dopaminergic treatment. Model evaluation was conducted using a strict leave-one-subject-out (LOSO) cross-validation protocol, ensuring that all recordings from a given participant were excluded from training when that subject was used for testing. To further assess the contribution and complementarity of these feature groups, models were trained using both the full feature set and systematically reduced feature sets generated through random ablation in which 50\% of features were removed. It is hypothesized that combinations of physiologically interpretable EEG features reflecting complementary aspects of oscillatory activity, network connectivity, and dynamical signal organization contain sufficient information to differentiate PD from healthy neural activity and to capture medication-related neural state changes under a rigorous subject-level validation framework.

\section*{Materials and Methods}

\subsection*{Participants} 
An openly available, previously published, resting-state EEG dataset was used for the analysis (openNeuro Accession Number ds002778). The dataset includes EEG data from 15 PD patients (eight female, mean age of 63.2 ± 8.2 years) on and off dopaminergic medication and 16 healthy, age-matched, control participants (nine female, mean age of 63.5 ± 9.6 years) and has been described extensively (George et al., 2013, Swann et al., 2015). All PD patients had been diagnosed by a movement disorder specialist at Scripps clinic in La Jolla, California. Participants provided written consent in accordance to the Institutional Review Board of the University of California, San Diego and the Declaration of Helsinki. For additional patient information the original study by George and colleagues (2013) can be consulted.

\subsection*{Data collection} 
Data from PD patients on and off medication were collected on different days with a counterbalanced order. For the on-medication recordings, patients continued their typical medication regimen. For the off-medication state, patients discontinued medication use at least 12 h before the session. Control participants were tested once. EEG data were acquired using a 32-channel BioSemi ActiveTwo system and were sampled at 512 Hz. Resting data were recorded for at least 3 min. During data collection, the participants were seated comfortably and told to fixate on a cross presented on a screen. Participants also completed several other assessments described in the previously published report (George et al.,2013), but these were not analyzed in the present report. 

\subsection*{EEG Data Preprocessing}
Raw EEG data (BioSemi .bdf format) were imported and preprocessed using the MNE-Python library (v1.11.0, \texttt{mne.io.read\_raw\_bdf}). Analyses were conducted using NumPy (v2.2.2) and SciPy (v1.13.1). After visual inspection of channel integrity, non-EEG external electrodes (EXG channels) were excluded from further analysis. Data were re-referenced to the common average reference using \texttt{mne.set\_eeg\_reference('average', projection=False)}. Continuous EEG signals were band-pass filtered between 0.5–45 Hz using a zero-phase finite impulse response (FIR) filter (\texttt{mne.filter(l\_freq=0.5, h\_freq=45, method='fir')}). The filter was applied using forward–reverse convolution to avoid phase distortion, equivalent to two-pass filtering as implemented in MATLAB’s \texttt{filtfilt}. Independent component analysis (ICA) was performed using the FastICA algorithm (\texttt{mne.preprocessing.ICA(method='fastica')}), which internally relies on the FastICA implementation from scikit-learn (v1.7.2). Prior to ICA decomposition, dimensionality reduction was performed via principal component analysis retaining 99.99\% of explained variance (\texttt{n\_components=0.999}). This procedure mirrors the extended Infomax/FastICA decomposition framework commonly used in EEGLAB, while operating within the MNE environment. Automatic component rejection was implemented based on three quantitative metrics computed from the ICA decomposition. Specifically, the projection power (sum of squared mixing matrix weights), kurtosis (via \texttt{scipy.stats.kurtosis}) and high-frequency muscle ratio (ratio of power at frequency range 25-45Hz over 1-15Hz) derived from Welch spectral estimates (\texttt{scipy.signal.welch}) were evaluated (Supplementary Figure S1). Projection power and kurtosis components exceeding percentile-based thresholds (95th percentile) and high frequency muscle ratio exceeding the threshold of three were marked for exclusion and removed during signal reconstruction with the accepted component only (\texttt{ica.apply()}). The cleaned continuous data were segmented into consecutive 5 s epochs with 1 s overlap, down-sampled to 256 Hz and stored as 3D NumPy arrays (epochs × channels × time points). Epochs exceeding ±15 standard deviations were rejected from further analysis. The EEG dataset preprocessing scripts are publicly accessible at the authors’ GitHub repository (\url{https://github.com/antoniosdougalis/Parkinson-s-Project.git}).

\subsection*{EEG feature extraction}

Preprocessed EEG epochs (NumPy arrays: epochs × 32 channels × 1280 samples) were subjected to a structured feature extraction pipeline routine implemented in Python (v3.10.18) using NumPy (v2.2.2), SciPy (v1.13.1), NeuroDSP (v2.3.0), PyBispectral (v1.3.0) and FOOOF (v1.1.1) libraries.

\subsection*{Time-domain statistics}

For each epoch and channel, mean, variance, and interquartile range (IQR = Q3-Q1) were computed using NumPy (\texttt{np.mean}, \texttt{np.var}, \texttt{np.percentile}) to quantify central tendency and signal dispersion. These measures quantify baseline amplitude characteristics and variability of ongoing neuronal activity. Changes in variance or dispersion may reflect altered cortical excitability or signal stability in pathological conditions.

\subsection*{Estimation of spectral aperiodic \& periodic components}

Power spectral density (PSD) was estimated using Welch’s method (\texttt{scipy.signal.welch}) using a Hann window, 50\% overlap and 0.5 Hz frequency resolution (\texttt{nperseg = 512} sample points, \texttt{nfft = 2*nperseg}). The PSD was subsequently utilized for computation of the aperiodic (offset, exponent) and aperiodic-corrected dominant peak periodic component using the FOOOF library as described previously (Donoghue et al., 2020). The model estimated, aperiodic offset and exponent, dominant spectral peak (center frequency [CF], bandwidth [BW], peak power [PW]) and goodness-of-fit (R²). Periodic peak threshold was set at 0.5dB and epochs with aperiodic fits with R² < 0.85 were rejected. Only a single dominant peak was allowed to be detected per epoch. The analysis separates periodic oscillatory activity from the aperiodic 1/f background component. The aperiodic exponent is thought to reflect synaptic excitation–inhibition balance (Gao et al., 2017), whereas oscillatory peaks represent structured rhythmic network activity.

\subsection*{Phase-based functional connectivity (PLV \& PLI)}
Narrowband phase time series were extracted using frequency-domain convolution with complex Morlet wavelets (\textit{fft}, \textit{ifft}, \textit{np.angle} via \textit{NumPy}/\textit{SciPy}) following Cohen’s method (2019). Four-second-long complex-valued Morlet wavelets (cMWs) were constructed by applying a Gaussian taper to a complex sine wave centered at a single frequency for each of the five canonical bands (central frequencies: delta, 2 Hz; theta, 6 Hz; alpha, 10 Hz; beta, 20 Hz; gamma, 38 Hz). The full width at half maximum (FWHM) of the Gaussian in the time domain (delta, 1.45 s; theta, 0.48 s; alpha, 0.25 s; beta, 0.22 s; gamma, 0.18 s) was selected to yield approximately 3–7 cycles per wavelet, enabling stable phase estimation and defining the effective frequency bandwidth of each filter (Supplementary Figure S2). Phase-locking value (PLV) and phase-lag index (PLI) between any pairs of electrodes across delta (1-4 Hz), theta (4-8 Hz), alpha (8-12 Hz), beta (13-30Hz) and gamma (30-45Hz) frequency bands were computed as proxies of inter-regional phase synchronization and connectivity.

PLV is defined as:
\[
\text{PLV}_{km} = \left| \frac{1}{N} \sum_{t=1}^{N} e^{j(\phi_k(t)-\phi_m(t))} \right|
\]

PLI is defined as:
\[
\text{PLI}_{km} = \left| \frac{1}{N} \sum_{t=1}^{N} \text{sign} \Big( \operatorname{Im} \{ e^{j(\phi_k(t)-\phi_m(t))} \} \Big) \right|
\]

where $\phi_k(t)$ and $\phi_m(t)$ are instantaneous phase time series (N sample points) of electrode $k$ and $m$, $j$ is the imaginary unit, and $\operatorname{Im}\{\cdot\}$ denotes the imaginary part. PLV and PLI are implemented in Python using the formulas PLV = abs(mean(exp(1j*$\Delta$phase))) and PLI = abs(mean(sign(imag(exp(1j*$\Delta$phase))))), where $\Delta$phase is the instantaneous phase difference between electrode pairs. PLV reflects consistency of phase differences between regions, indexing functional coupling. PLI reduces sensitivity to zero and $2\pi$ radians-lag synchronization that often reflect volume conduction effects at the expense of losing information on true zero/$2\pi$ lag interactions.

\subsection*{Surface Laplacian transform}
To evaluate whether phase synchronization effects could be influenced by spatial spread of scalp potentials, an additional analysis was performed on Laplacian-transformed EEG signals. The surface Laplacian provides an estimate of the second spatial derivative of the scalp potential distribution and acts as a spatial high-pass filter that attenuates broadly distributed fields arising from volume conduction while emphasizing more local cortical activity (Supplementary Figure S3). The transform was computed using the spherical spline interpolation method of Perrin and colleagues (Perrin et al., 1989) implemented in Python following the modified formulation described by Cohen (2014a). Electrode coordinates were projected onto a unit sphere and Legendre polynomial expansions were used to construct the spline weighting matrices that map scalp potentials to current source density estimates. A small smoothing parameter ($10^{-5}$) was applied to stabilize the inversion of the spline matrix.

The resulting Laplacian-transformed signals were then used to recompute phase-based connectivity metrics (PLV and PLI) using the same procedures described above. In addition, for these transformed signals only, the weighted Phase Lag Index (wPLI) was computed as:
\[
\text{wPLI}_{km} = 
\frac{\left| \mathbb{E}\left[ \, \big| \operatorname{Im}\{ e^{i(\phi_k(t)-\phi_m(t))} \} \big| \cdot 
\text{sign} \big( \operatorname{Im}\{ e^{i(\phi_k(t)-\phi_m(t))} \} \big) \, \right] \right|}
{\mathbb{E}\left[ \, \big| \operatorname{Im}\{ e^{i(\phi_k(t)-\phi_m(t))} \} \big| \, \right]}
\]

where $\phi_k(t)$ and $\phi_m(t)$ are the instantaneous phase time series of electrodes $k$ and $m$, $i$ is the imaginary unit, $\operatorname{Im}\{\cdot\}$ extracts the imaginary part, $\text{sign}(\cdot)$ is the sign function, and $\mathbb{E}[\cdot]$ denotes the expectation value over time (sample mean). wPLI was computed in Python as wpli = np.abs(np.mean(np.sin($\Delta$phase))) / np.mean(np.abs(np.sin($\Delta$phase))), where $\Delta$phase is the instantaneous phase difference between electrode pairs. This analysis was performed only on Laplacian-transformed data to assess whether the observed PLV and PLI effects could be explained by volume conduction. The wPLI was not computed for the untransformed EEG, and its inclusion here allows a more robust evaluation of true near-zero-lag versus lagged phase synchrony after spatial leakage was reduced.

\subsection*{Long-range temporal correlations (LRTC) via Detrended Fluctuation Analysis (DFA)}
Detrended fluctuation analysis was performed using SciPy and NeuroDSP as described previously (Cole et al., 2019; Hardstone et al., 2012). First, narrowband signals were extracted using a 3rd-order Butterworth bandpass filter (scipy.signal.butter, filtfilt). The amplitude envelope was then computed via the Hilbert transform (scipy.signal.hilbert) and DFA was computed across ten logarithmically spaced window sizes (window time length min-max, 2–90 s). DFA exponents were computed for canonical frequency bands: delta (1-4 Hz), theta (4-8 Hz), alpha (8-12 Hz), beta (13-30Hz) and gamma (30-45Hz). The DFA exponent reflects scale-free persistence in neuronal dynamics, often interpreted as proximity to critical network states (Hardstone et al., 2012).

\subsection*{Functional excitation–inhibition ratio (fE/I)}
The fE/I metric (Bruining et al., 2020) was implemented in Python based on publicly available MATLAB code. Narrowband filtering was performed via Gaussian frequency-domain filtering at canonical frequency bands. The amplitude normalized detrended signal profile and normalized fluctuation function were computed per overlapping 5 s windows (80\% overlap) and fE/I was computed as:
\[
\text{fE/I} = 1 - \text{Pearson's correlation}(\text{amplitude envelope}, \text{normalized fluctuation function})
\]

The fE/I metric estimates the dynamic balance between excitatory and inhibitory processes by relating amplitude envelopes to normalized fluctuation functions. Deviations may indicate dysregulated cortical gain control mechanisms.

\subsection*{Phase–amplitude coupling (PAC)}
Cross-frequency coupling was estimated using antisymmetrized-corrected, normalized bicoherence implemented via PyBispectral as described previously (Binns et al., 2025). PAC strength was defined as the magnitude of normalized bispectral coefficients between low-frequency phase and high-frequency amplitude bands. PAC measures whether slow oscillatory phase modulates high-frequency amplitude, reflecting hierarchical cross-frequency communication. This mechanism is central to coordinating local processing (gamma) within slower large-scale network rhythms.

\subsection*{Neuronal avalanche and critical dynamics analysis}
Neuronal avalanches were detected from broadband EEG signal for every epoch via amplitude threshold crossings (5\% of maximum epoch amplitude) based on previously described methods (Bruining et al., 2020; Shew et al., 2009). Avalanche size and duration were computed per electrode. The $\kappa$ index of criticality was estimated by comparing empirical cumulative avalanche size distributions to a reference power-law (exponent $-1.5$) across ten logarithmically spaced bins (from min to max avalanche size). Avalanche analysis assesses whether neuronal activity follows scale-free power-law distributions consistent with criticality. The $\kappa$ index quantifies deviation from theoretical critical dynamics.

\subsection*{Instantaneous frequency (IF) analysis (frequency sliding)}
Temporal fluctuations in oscillatory frequency were quantified using the frequency sliding method described by Cohen (2014b). Narrowband filtered signals were obtained via frequency-domain convolution with appropriate filter kernels at canonical frequencies and IF was computed from the temporal derivative of the unwrapped Hilbert phase:
\[
f(t) = f_s \cdot \frac{d}{dt} \left( \frac{\text{unwrap}(\angle(H[y(t)]))}{2\pi} \right)
\]

where \(y\) is the bandpass-filtered signal, $\mathcal{H}$ denotes the Hilbert transform, \texttt{unwrap} removes the circular discontinuities in the phase and presents it as a continuous function, \(\angle\) represents the angle of the phase (i.e., the phase at each instance), and \(f_s\) is the sampling frequency. IF time series were evaluated at 125 time points per 5-second epoch (effective sampling rate 25 Hz). The function is implemented in Python as \textit{fs* (np.diff( np.unwrap( np.angle( hilbert(y) ) ))/(2*np.pi))}. Multiscale median smoothing in ten windows covering 10 to 100 sample points was applied as described previously to smooth the resultant IF time series from abrupt off-range angle difference values (Cohen, 2014b). Amplitude-weighted frequency estimates were computed to reduce bias from low-amplitude segments. Inter-electrode coordination was quantified using pairwise Spearman correlations of instantaneous frequency time series. This procedure yields a square correlation matrix describing inter-electrode coordination of instantaneous frequency fluctuations. 

To characterize dynamic frequency behavior, three additional quantitative descriptors were extracted per epoch, electrode, and frequency band. First, the range of IF fluctuations was computed as \( \text{IF}_{\text{range}} = P_{95}(f(t)) - P_5(f(t)) \), using \textit{np.percentile}, indexing the amplitude of frequency excursions. Second, the temporal smoothness (frequency jitter index) was computed via \textit{np.diff} and \textit{np.nanmean} as \textit{IF\_diff = "mean"(|f(t)-f(t-1)|)}, quantifying short-timescale variability of frequency changes. Last, the second-order modulation frequency (frequency of the frequency), representing the dominant modulation frequency of the instantaneous frequency time series, was extracted from the peak of its Fourier spectrum (\textit{np.fft.rfft, np.argmax}), excluding the lowest frequencies. This measure captures rhythmic structure in frequency fluctuations themselves. 

The frequency sliding method captures moment-to-moment fluctuations in oscillatory frequency rather than static band power. Coordinated frequency fluctuations across electrodes may reflect dynamic network reconfiguration. Instantaneous frequency reflects moment-to-moment changes in network excitability and oscillatory pacing. The range indexes flexibility of oscillatory generators, the jitter index captures temporal stability versus instability, and the modulation frequency quantifies higher-order rhythmic organization of frequency dynamics.

\subsection*{Harmonic frequency ratio locking analysis (n:m instantaneous frequency locking)}
Harmonic locking between oscillatory bands was quantified from instantaneous frequency time series derived via frequency sliding. The approach follows the framework proposed by Scheffer-Teixeira \& Tort (2016), in which integer frequency ratios are interpreted as a mechanism facilitating cross-frequency coordination. For each epoch and electrode, instantaneous frequency ratios were computed as:

\[
R(t) = \frac{f_{\text{high}}(t)}{f_{\text{low}}(t)}
\]

Two harmonic relationships were evaluated at alpha/theta (2:1) and gamma/theta (5:1). Ratios were rounded to one decimal precision (np.round) and time points were classified as harmonically locked when R(t) equaled the expected integer ratio (2.0 or 5.0). Binary masks were segmented into contiguous events using connected-component labeling (\textit{scipy.ndimage.label}). Although event counts and total durations were computed, the primary outcome measure used in statistical analyses was the percentage of epoch time spent in harmonic locking, defined as:
\[
\%HL = 100 \times \frac{\text{Number of locked samples}}{\text{Total samples}}
\]
Integer frequency relationships are proposed to enable stable phase alignment across oscillatory bands, facilitating inter-frequency communication and dynamic integration of neuronal processes. Increased harmonic locking reflects stronger hierarchical coordination between slow and fast rhythms.

\subsection*{EEG feature sets for Deep Learning (DL) classification}
Resting-state EEG recordings were processed using the multi-stage feature extraction pipeline designed to capture oscillatory, scale-free, and network-level dynamics. For each subject, features were extracted independently from 32 scalp electrodes per epoch and were average per subject forming a three-dimensional tensor of subjects × electrodes × features. Connectivity measures were summarized per source electrode by averaging connectivity of the square matrix with all other target electrodes. This preserves global integration properties while reducing dimensionality and maintaining compatibility with token-based transformer modelling. Seventy-seven features were computed, including time-domain statistics, relative and absolute spectral power, phase-based connectivity metrics, IF measures, aperiodic exponent and periodic-corrected dominant peak characteristics, DFA, functional excitation–inhibition ratio (fEI), phase–amplitude coupling, neuronal avalanche metrics (size and κ) and harmonic instantaneous frequency locking. To reduce redundancy and improve interpretability of the feature sets for deep learning classification, features exhibiting extremely high correlations were removed prior to final set formation. Specifically, pairwise correlations were computed across all subjects, and features with correlations approaching $\rho \approx 1$ were considered redundant and excluded. For example, in the time domain, interquartile range and variance were highly correlated, while many phase–amplitude coupling and phase–phase connectivity ratios similarly showed near-perfect correlations. Following this pre-screening, 40 features were retained: 17 for the Standard set (time-domain, spectral power, and phase connectivity measures) and 23 for the Dynamical set (aperiodic, scale-free, cross-frequency, neuronal avalanche, and instantaneous frequency descriptors). This procedure ensures that the final sets capture largely complementary aspects of EEG dynamics while minimizing multicollinearity, thereby improving the interpretability and stability of downstream deep learning models. 

\subsection*{Handling of missing feature values}
Occasional missing values due to instability in feature estimation were imputed independently per electrode–feature pair using local interpolation across subjects. Missing entries were replaced by the mean of up to four neighboring subjects (two preceding, two subsequent). Boundary cases used all available neighbors. This preserved electrode-specific feature distributions while minimizing distortion of inter-subject variability.

\subsection*{Feature normalization}
Features were standardized via z-score normalization independently for each electrode–feature pair. Means and standard deviations were computed across subjects and applied to transform features to zero mean and unit variance, preserving spatial specificity while ensuring comparable numerical scales.

\subsection*{Multi-Head Attention Transformer-based classification model}
All DL activities, including construction of the multi-head attention transformer model and running of training, were executed in Python using the \textit{PyTorch} library (v2.4.1) either locally, on a personal computer equipped with an AMD Ryzen 5 PRO processor and 16 GB RAM, or in Google Colab using CUDA-enabled GPU acceleration when required. Membership classification was performed using a custom transformer encoder architecture with multi-head self-attention, based on the original description by Vaswani and colleagues (Vaswani \textit{et al.,} 2017). Each electrode was treated as a token represented by its corresponding feature vector. Electrode feature vectors were projected into a shared embedding space using a linear embedding layer. The encoder consisted of six stacked transformer blocks, each comprising multi-head self-attention and a feed-forward multilayer perceptron. Each transformer block contained a feed-forward multilayer perceptron with hidden dimensionality of 256 units. Layer normalization preceded each submodule and residual connections were used throughout, following the pre-normalization transformer architecture described by Vaswani \textit{et al.} (2017). Multi-head self-attention employed six parallel heads operating on 60-dimensional embeddings (10 dimensions per head). Following encoding, electrode embeddings were aggregated via mean pooling and passed through layer normalization and a final linear classification layer mapping the 60-dimensional pooled embedding to three output classes: control (CN), PD off medication (PDoff), and PD on medication (PDon). Dropout regularization (rate = 0.1) was applied throughout the encoder layers. The full implementation of the transformer model and training procedure is publicly available at the authors’ GitHub repository ( \url{https://github.com/antoniosdougalis/Parkinson-s-Project.git} ). 

\subsection*{Model training and leave-one-subject-out (LOSO) cross-validation strategy}
Data were presented to the transformer model in batches of six subjects during training. A Cross-entropy loss function from the PyTorch library was used for raw logits classification (\texttt{nn.CrossEntropyLoss()}). An Adam optimizer was used with a learning rate of 0.0001. An adaptive learning-rate scheduler ((\texttt{torch.optim.lr\_scheduler.ReduceLROnPlateau})) was employed to help the model escape local minima. The scheduler halved the learning rate after two consecutive epochs without improvement in loss (threshold = 0.01). Model performance was evaluated using leave-one-subject-out cross-validation (LOSO). For each fold, one subject was held out for testing and the remaining subjects used for training. Normalization parameters were computed exclusively on training data to prevent information leakage. Fifty training epochs were employed for each subject for both two-class and three-class classification runs. Training was stopped early when both conditions loss $< 0.03$ and accuracy $> 99.7\%$ were met.

\subsection*{Generation of random reduced feature sets}
To evaluate feature importance, random subsets containing 50\% of the features were generated (\texttt{np.random.choice}). Classification was performed on each reduced set and mean accuracy of the random sets was compared to that of the full set. One hundred random reductions were performed for each group contrast comparison for each feature set in order to keep computational time to a minimum.

\subsection*{Statistical analysis}
\subsection*{Assessing model performance }
All analyses were conducted in Python using scikit-learn (v1.6.1) and statsmodels (v0.14.5). Confusion matrices (\texttt{sklearn.metrics.confusion\_matrix}) were computed per condition. Using True Positive (TP), False Positive (FP), True Negative (TN) and False Negative (FN), the following metrics were calculated: Accuracy = (TP + TN) / (TP + TN + FP + FN);  Recall (Sensitivity) = TP / (TP + FN);  Precision = TP / (TP + FP); F1 = 2 × (Precision × Recall) / (Precision + Recall) ; Specificity = TN / (TN + FP). Macro-averaged metrics were computed across classes. Balanced accuracy was defined as mean recall, and macro-F1 as mean class-wise F1. ROC--AUC (\texttt{roc\_auc\_score from sklearn.metrics}) was computed using predicted probabilities. Binary AUC used the positive class probability. Multiclass AUC employed a weighted one-vs-rest approach (\texttt{multi\_class='ovr', average='weighted'}).

\subsection*{Statistical comparisons of model performance \& Confidence Intervals (CI)}
Performance differences between models utilizing different sets of features were evaluated using McNemar’s test (\texttt{statsmodels.stats.contingency\_tables.mcnemar, exact=True}). Subject-level 2×2 contingency tables were constructed from paired correctness labels of the two models being compared. The McNemar’s test evaluates symmetry of discordant pairs (b = c under $H_0$)), testing whether misclassifications favor one model over the other. CIs of model accuracy were estimated via non-parametric bootstrapping (\texttt{sklearn.utils.resample}, 5000 iterations, resampled with replacement). For each resample, accuracy was recomputed and 95\% CIs were derived from the 2.5th and 97.5th percentiles (\texttt{np.percentile}). Similarly, paired subject-level accuracy differences of model A and B were computed as:

\[
\text{diff}_i = \text{correct}_{A,i} - \text{correct}_{B,i}
\]

where i indexes each subject. Mean ΔAccuracy represents overall performance difference between two model conditions with a paired bootstrap procedure (5000 iterations, resampled with replacement) providing the 95\% CIs.

\subsection*{Statistical analysis of the random feature ablations}
To evaluate whether the full feature sets carry more predictive information than random subsets, the accuracy of the full model was compared to one hundred models trained on randomly reduced feature sets (50\% of features removed). For each condition and feature set, the empirical p-values were calculated as the proportion of random models that achieved accuracy greater than or equal to the full model:

\[
p = \frac{\left(\#\,\text{random accuracies} \ge \text{full model accuracy}\right) + 1}
{\left(\text{number of random permutations}\right) + 1}
\]

\subsection*{Feature correlation analysis}
Spearman correlation coefficient and corresponding p values were computed for all pairs of features across all subjects (\texttt{spearmanr, from scipy.stats}).  Multiple comparisons were controlled using false discovery rate (FDR) correction (\texttt{fdrcorrection, from statsmodels.stats.multitest}) using the Benjamini/Hochberg method. Additionally, 95\% CI of the correlation coefficient were estimated via bootstrapping (5000 resamples) for each pairwise comparison. Correlation matrices were visualized with lower-triangle coefficients and upper-triangle confidence intervals, with significance maps displayed separately. A redundancy index was computed for each feature set as the mean absolute pairwise correlation coefficient (mean |ρ|) across all unique feature pairs.

\subsection*{Statistical analysis of group comparisons}
Group differences in EEG-derived features were evaluated using non-parametric statistical tests. For each feature, subject-level values were obtained by averaging across electrodes (either across the full electrode set or within the sensorimotor region of interest). Pairwise comparisons were then performed between the three study groups: healthy controls (CN), Parkinson’s disease patients off medication (PDoff), and the same patients on medication (PDon). Comparisons between independent groups (CN vs PDoff and CN vs PDon) were assessed using the Wilcoxon rank-sum test whereas comparisons between medication states within patients (PDoff vs PDon) were assessed using the Wilcoxon signed-rank test for paired samples. To control for multiple comparisons, false discovery rate (FDR) correction was applied within predefined families of related features (e.g., across frequency bands within the same connectivity metric). Effect sizes for pairwise comparisons were quantified using Cliff’s delta (δ) for independent samples (CN vs PD comparisons) and the matched-pairs rank-biserial correlation for paired comparisons (PDoff vs PDon), with confidence intervals estimated via bootstrap resampling (5000 iterations). This approach provides both statistical significance and an estimate of the magnitude and reliability of observed effects.

\section*{Results}

\subsection*{Classification Performance and Between--Feature Set Comparisons}

Overall classification performance across diagnostic contrasts is summarized in Table 1 and illustrated in Figure 1, with corresponding ROC curves shown in Figure 2. Prior to model training, features were pre-screened only to remove highly redundant measures, resulting in 17 Standard and 23 Dynamical features that capture complementary aspects of EEG dynamics, ensuring that subsequent classification performance reflects distinct information rather than redundancy. Models were trained using the Standard and Dynamical EEG feature sets, and a Fusion configuration combining their decision outputs. Direct comparisons between feature configurations are presented in Table 2. Across most diagnostic contrasts, differences between feature sets did not reach statistical significance after FDR correction, indicating that neither the Standard nor the Dynamical feature configuration consistently outperformed the other across all classification tasks. Nevertheless, the pattern of results differed depending on the diagnostic contrast. The clearest effect was observed in the PD\textsubscript{off} vs PD\textsubscript{on} comparison, where the Standard feature set significantly outperformed the Dynamical set, as indicated by McNemar’s test after FDR correction. In the same contrast, the Standard configuration also significantly exceeded the Fusion model. This suggests that features capturing traditional spectral power and synchronization properties may be particularly informative for distinguishing medication states within Parkinson’s disease. In contrast, no statistically significant differences were observed between configurations in the three-class classification or in the CN--PD\textsubscript{off} and CN--PD\textsubscript{on} contrasts, despite differences in observed accuracy values. In these tasks, the Dynamical feature set often showed numerically higher performance for contrasts involving cognitively normal participants and Parkinson’s disease patients, although these differences did not reach statistical significance.

\begin{figure}[h!]
	\centering
	\includegraphics[width=1.0\textwidth]{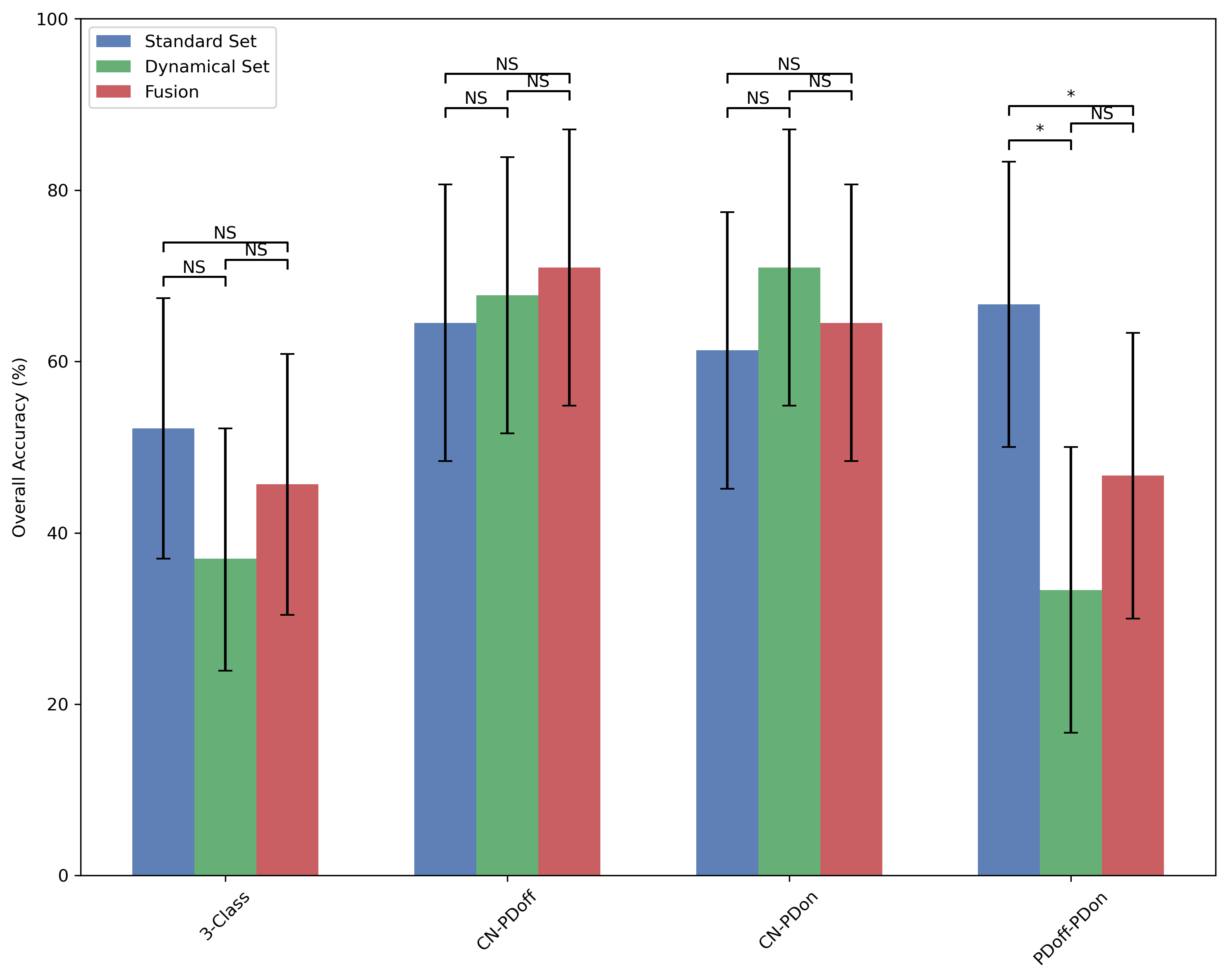}
	\captionsetup{justification=justified}
	\caption{ Classification accuracy across diagnostic conditions using EEG feature configurations in a multi-head attention transformer model. Performance is shown for four tasks: three-class classification (CN, PD\textsubscript{off}, PD\textsubscript{on}) and pairwise contrasts (CN--PD\textsubscript{off}, CN--PD\textsubscript{on}, PD\textsubscript{off}--PD\textsubscript{on}). Models were trained separately on the Standard and Dynamical feature sets using strict leave-one-subject-out (LOSO) cross-validation. Fusion corresponds to the decision-level model output obtained by averaging the softmax probabilities of the Standard and Dynamical feature sets. Bars indicate mean accuracy with 95\% bootstrap confidence intervals (5000 samples). Statistical comparisons were performed using McNemar’s test on contingency tables of classification correctness with FDR correction (*p $<$ 0.05; NS, not significant).}
\end{figure}

\begin{figure}[h!]
	\centering
	\includegraphics[width=1.0\textwidth]{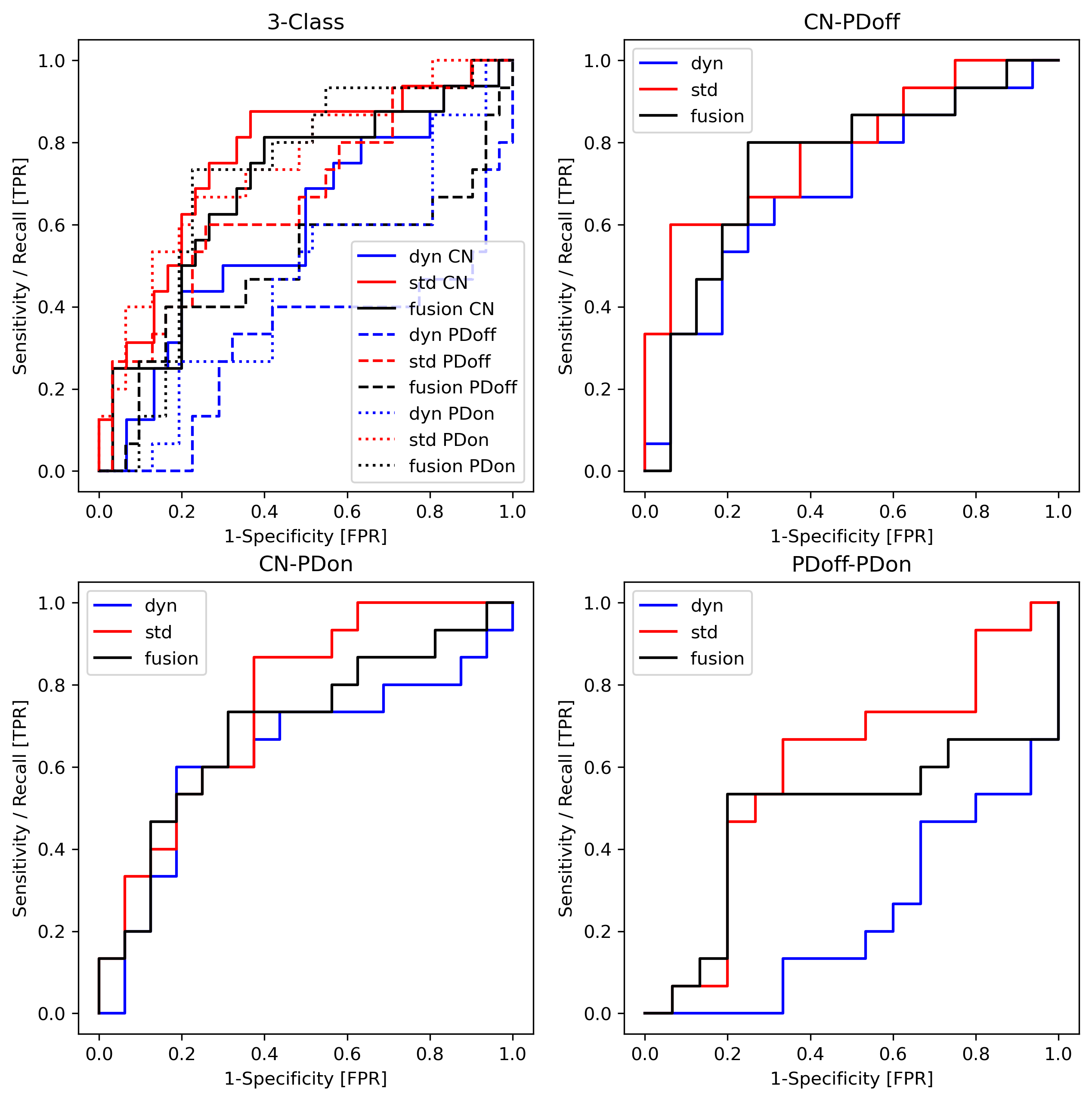}
	\captionsetup{justification=justified}
	\caption{ Receiver operating characteristic (ROC) curves for transformer models across diagnostic conditions. Curves are shown for models trained on the Standard and Dynamical feature sets and for Fusion, which corresponds to the decision-level model output obtained by averaging the softmax probabilities of the Standard and Dynamical feature sets. For the three-class problem, ROC curves were computed using a one-vs-rest strategy for each group (CN, PD\textsubscript{off}, PD\textsubscript{on}). Performance is summarized by the weighted ROC--AUC (TPR, True Positive Rate; FPR, False Positive Rate).}
\end{figure}

\begin{table}[htbp]
	\centering
	\renewcommand{\arraystretch}{1.3}
	
	\begin{tabularx}{\textwidth}{>{\centering\arraybackslash}X *{4}{>{\centering\arraybackslash}X}}
		\toprule
		
		\multicolumn{5}{c}{\textbf{Standard Features}} \\
		\midrule
		& \textbf{CN vs PDoff vs PDon} & \textbf{CN vs PDoff} & \textbf{CN vs PDon} & \textbf{PDoff vs PDon} \\
		\midrule
		Overall Accuracy [CI low, high] & 52.17 [39.13, 67.39] & 64.52 [48.39, 80.65] & 61.29 [45.16, 77.42] & 66.67 [50.0, 83.33] \\
		Balanced Accuracy & 52.34 & 64.58 & 61.25 & 66.67 \\
		Macro F1 & 52.38 & 64.52 & 61.25 & 66.67 \\
		ROC-AUC & 72.52 & 78.75 & 75.83 & 59.56 \\
		
		\midrule
		\multicolumn{5}{c}{\textbf{Dynamical Features}} \\
		\midrule
		Overall Accuracy [CI low, high] & 36.96 [23.91, 50.0] & 67.74 [51.61, 83.87] & 70.97 [54.84, 87.1] & 33.33 [16.67, 50.0] \\
		Balanced Accuracy & 36.81 & 67.5 & 70.62 & 33.33 \\
		Macro F1 & 37.08 & 67.44 & 70.48 & 33.33 \\
		ROC-AUC & 45.96 & 68.75 & 63.33 & 23.56 \\
		
		\midrule
		\multicolumn{5}{c}{\textbf{Fusion model output}} \\
		\midrule
		Overall Accuracy [CI low, high] & 45.65 [30.43, 60.87] & 70.97 [54.84, 87.1] & 64.52 [48.39, 80.65] & 46.67 [30.0, 63.33] \\
		Balanced Accuracy & 45.56 & 70.83 & 64.17 & 46.67 \\
		Macro F1 & 46.13 & 70.85 & 63.92 & 46.43 \\
		ROC-AUC & 63.25 & 74.58 & 69.58 & 48.0 \\
		
		\bottomrule
	\end{tabularx}
	
	\caption{Classification performance across diagnostic contrasts (3-Class, CN--PD\textsubscript{off}, CN--PD\textsubscript{on}, PD\textsubscript{off}--PD\textsubscript{on}) using the Standard and Dynamical feature sets and Fusion. Fusion corresponds to the model output obtained by averaging the softmax probabilities of the Standard and Dynamical feature sets. Overall accuracy is reported with 95\% bootstrap confidence intervals, along with balanced accuracy, macro-F1 score, and weighted ROC--AUC.}

\end{table}

\begin{table}[htbp]
	\centering
	\scriptsize
	\renewcommand{\arraystretch}{1.3}
	\resizebox{\textwidth}{!}{%
		\begin{tabularx}{\textwidth}{>{\centering\arraybackslash}X *{3}{>{\centering\arraybackslash}X}}
			\toprule
			
			\multicolumn{4}{c}{\textbf{\% $\Delta$Acc, $\Delta$Acc CI [low, high], p-McNemar’s (FDR-corrected)}} \\
			\midrule
			\multicolumn{4}{c}{\textbf{Standard vs. Fusion}} \\ 
			\midrule
			\textbf{CN vs PDoff vs PDon} & \textbf{CN vs PDoff} & \textbf{CN vs PDon} & \textbf{PDoff vs PDon} \\ 
			\makecell[c]{6.52, [-6.52, 19.57] \\ 0.508 (0.508)} &
			\makecell[c]{-6.45, [-19.35, 6.45] \\ 0.625 (1)} &
			\makecell[c]{-3.23, [-19.35, 9.68] \\ 1 (1)} &
			\makecell[c]{20.0, [6.67, 33.33] \\ 0.031 (0.047)} \\ 
			
			\midrule
			\multicolumn{4}{c}{\textbf{Dynamical vs. Fusion}} \\
			\midrule
			\textbf{CN vs PDoff vs PDon} & \textbf{CN vs PDoff} & \textbf{CN vs PDon} & \textbf{PDoff vs PDon} \\
			\makecell[c]{-8.7, [-21.74, 4.35] \\ 0.344 (0.508)} &
			\makecell[c]{-3.23, [-16.13, 9.68] \\ 1 (1)} &
			\makecell[c]{6.45, [-6.45, 19.35] \\ 0.625 (0.938)} &
			\makecell[c]{-13.33, [-30.0, 3.33] \\ 0.289 (0.289)} \\
			
			\midrule
			\multicolumn{4}{c}{\textbf{Dynamical vs. Standard Features}} \\
			\midrule
			\textbf{CN vs PDoff vs PDon} & \textbf{CN vs PDoff} & \textbf{CN vs PDon} & \textbf{PDoff vs PDon} \\
			\makecell[c]{-15.22, [-32.61, 2.17] \\ 0.167 (0.501)} &
			\makecell[c]{3.23, [-16.13, 22.58] \\ 1 (1)} &
			\makecell[c]{9.68, [-9.68, 29.03] \\ 0.508 (0.938)} &
			\makecell[c]{-33.33, [-53.3, -13.33] \\ 0.013 (0.039)} \\
			\bottomrule
		\end{tabularx}%
	}
	
	\caption{Pairwise classification accuracy differences between EEG configurations across diagnostic contrasts. Comparisons are shown for Dynamical vs Fusion, Standard vs Fusion, and Dynamical vs Standard. Fusion corresponds to the model output obtained by averaging the softmax probabilities of the Standard and Dynamical feature sets. Values report mean \% $\Delta$Accuracy, 95\% bootstrap confidence intervals, and McNemar’s test p-values (FDR-corrected). Positive $\Delta$Accuracy indicates higher accuracy for the first feature set listed in each comparison.}
	
\end{table}

Overall, these findings indicate that the relative contribution of feature configurations depends on the specific diagnostic contrast, with the Standard set providing a clear advantage for discriminating medication states within Parkinson’s disease, while the Dynamical features show competitive (or numerically stronger) performance to the Standard Set in contrasts involving healthy controls.

\subsection*{Comparison of Full Feature Sets with Random Ablations}

To further evaluate whether model performance depended on the full feature configurations, classification accuracies were compared with models trained on randomly reduced feature subsets, in which 50\% of the features were removed. These results are summarized in Table 3 and visualized in Figure 3. For contrasts involving cognitively normal participants and Parkinson’s disease patients (CN--PD\textsubscript{off} and CN--PD\textsubscript{on}), the Dynamical feature set consistently performed near the upper range of the random-ablation distributions, with the full model accuracy exceeding the mean accuracy obtained from the randomly reduced feature sets. In contrast, the Standard feature set showed little separation from the random-ablation distributions, with the full model accuracy lying close to the middle or lower portion of the random accuracy range. This pattern is consistent with the interpretation that the predictive performance of the Dynamical feature configuration in these contrasts depends on the joint contribution of multiple complementary features, such that removing random subsets of features tends to degrade classification performance. In contrast, the Standard feature configuration appears less sensitive to random feature removal in these contrasts, indicating weaker complementarity between individual features. A different pattern was observed in the PD\textsubscript{off} vs PD\textsubscript{on} contrast, where the Standard feature set showed clearer separation from the random-ablation distributions, whereas the Dynamical configuration performed similarly to its random subsets. This finding mirrors the statistical comparison results described above, where the Standard feature set significantly outperformed the Dynamical configuration in this task. In the 3-class contrast, both the Standard and the Dynamical feature configurations sets showed good separation from the random-ablation distributions.

\begin{figure}[h!]
	\centering
	\includegraphics[width=1.0\textwidth]{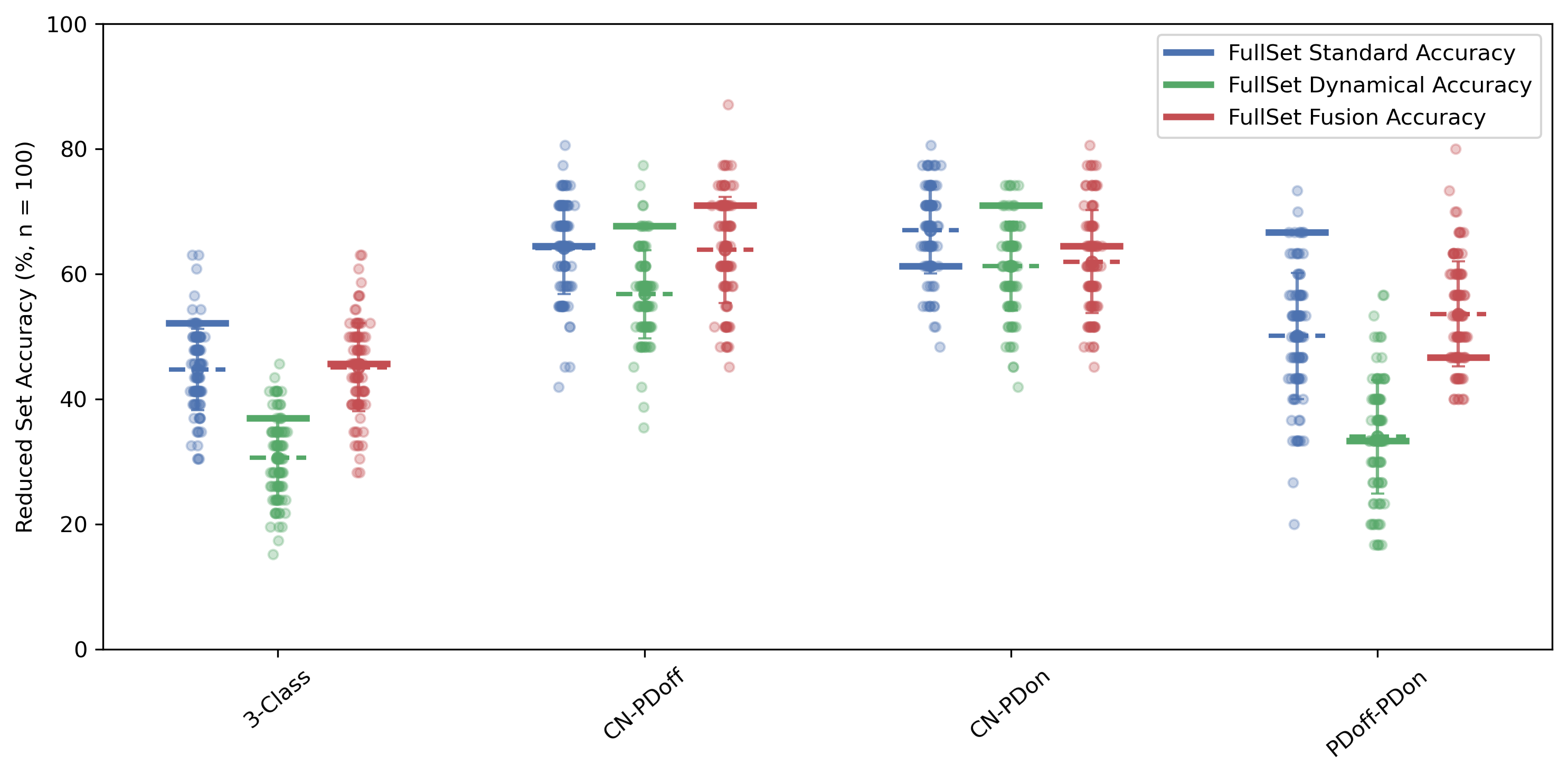}
	\captionsetup{justification=justified}
	\caption{ Accuracy of EEG feature configurations compared with random feature reductions across diagnostic contrasts (3-Class, CN--PD\textsubscript{off}, CN--PD\textsubscript{on}, PD\textsubscript{off}--PD\textsubscript{on}) using the Standard and Dynamical feature sets and Fusion. Fusion corresponds to the decision-level output obtained by averaging the softmax probabilities of the Standard and Dynamical sets. For each feature set, the solid horizontal line indicates the accuracy of the full model. Scatter points represent accuracies from one hundred models trained on randomly reduced feature subsets (50\% features removed). The dashed horizontal line indicates the mean accuracy of the random ablations while the vertical lines represent its standard deviation.}
\end{figure}

\begin{table}[htbp]
	\centering
	\renewcommand{\arraystretch}{1.3}
	
	\begin{tabularx}{\textwidth}{>{\centering\arraybackslash}X *{4}{>{\centering\arraybackslash}X}}
		\toprule
		
		\multicolumn{5}{c}{\textbf{Full Set Accuracy, Random Set Accuracy (mean $\pm$ St.D), emp.p-val}} \\
		
		\midrule
		& \textbf{CN vs PDoff vs PDon} & \textbf{CN vs PDoff} & \textbf{CN vs PDon} & \textbf{PDoff vs PDon} \\
		\midrule
		
		\textbf{Standard Features} 
		& 52.17, (44.72 $\pm$ 6.52), 0.129 
		& 64.52, (64.19 $\pm$ 7.42), 0.624 
		& 61.29, (67.0 $\pm$ 6.93), 0.861 
		& 66.67, (50.1 $\pm$ 10.12), 0.089 \\
		
		\textbf{Dynamical Features} 
		& 36.96, (30.63 $\pm$ 6.5), 0.198 
		& 67.74, (56.77 $\pm$ 7.07), 0.109 
		& 70.97, (61.26 $\pm$ 7.1), 0.119 
		& 33.33, (34.0 $\pm$ 9.07), 0.663 \\
		
		\textbf{Fusion} 
		& 45.65, (45.09 $\pm$ 7.05), 0.564 
		& 70.97, (63.87 $\pm$ 8.49), 0.317 
		& 64.52, (62.0 $\pm$ 8.23), 0.436 
		& 46.67, (53.63 $\pm$ 8.42), 0.842 \\
		
		\bottomrule
	\end{tabularx}
	
	\caption{Comparison of full feature configurations with randomly reduced feature sets across diagnostic contrasts (3-Class, CN--PDoff, CN--PDon, PDoff--PDon). For each feature set (Standard, Dynamical) and Fusion (decision-level output obtained by averaging the softmax probabilities of Standard and Dynamical sets), the table reports the accuracy of the full model, the mean $\pm$ SD accuracy from one hundred random feature reductions (50\% features removed) and the empirical permutation p-value, defined as the proportion of random models achieving accuracy greater than or equal to the full model.}
	
\end{table}

Taken together, the ablation analysis suggests that the informational structure of the two feature configurations differs. In contrasts involving Parkinson’s disease versus controls, the Dynamical feature set appears to rely on complementary information distributed across features, such that random feature removal reduces classification performance. In contrast, the Standard feature set appears less dependent on such complementary interactions between features in these tasks, while showing greater sensitivity in the medication-state contrast.

\subsection*{Correlations Between Features: A case of non-redundancy}

To assess potential redundancy among features within each configuration, pairwise correlations were computed across all subjects for both the Standard and Dynamical feature sets (Figures 4 and 5). Within the Standard feature set, pairwise correlations were generally moderate, with the largest observed correlations occurring between $\alpha$-band synchronization and power measures (e.g., $\alpha$-PLV vs $\alpha$ relative power, $\rho \approx 0.77$; $\alpha$-PLI vs $\alpha$ relative power, $\rho \approx 0.76$). Most feature pairs exhibited correlations well below $|0.5|$, and the overall redundancy index (mean $|\rho| = 0.263$) indicates limited overlap among descriptors derived from different spectral bands or methodological families. In the Dynamical feature set, the majority of feature pairs also showed modest correlations. A few isolated pairs exhibited higher correlations, such as $\alpha$-band instantaneous frequency range vs $\alpha$-band instantaneous frequency difference ($\rho \approx 0.95$) and neuronal avalanche size vs aperiodic spectral offset ($\rho \approx 0.88$). Aside from these exceptions, most pairwise associations remained modest, consistent with a low overall redundancy index (mean $|\rho| = 0.260$), indicating that the features capture largely complementary aspects of neural dynamics. Across all pairwise comparisons, 26 of 136 unique feature pairs ($\approx 19\%$) in the Standard feature set and 47 of 253 pairs ($\approx 19\%$) in the Dynamical feature set showed statistically significant correlations after FDR correction. Thus, although some relationships reached statistical significance, the majority of pairwise associations remained modest, supporting the interpretation that the included features capture distinct and complementary aspects of EEG dynamics rather than redundant measurements of the same signal properties.

\begin{figure}[h!]
	\centering
	\includegraphics[width=1.0\textwidth]{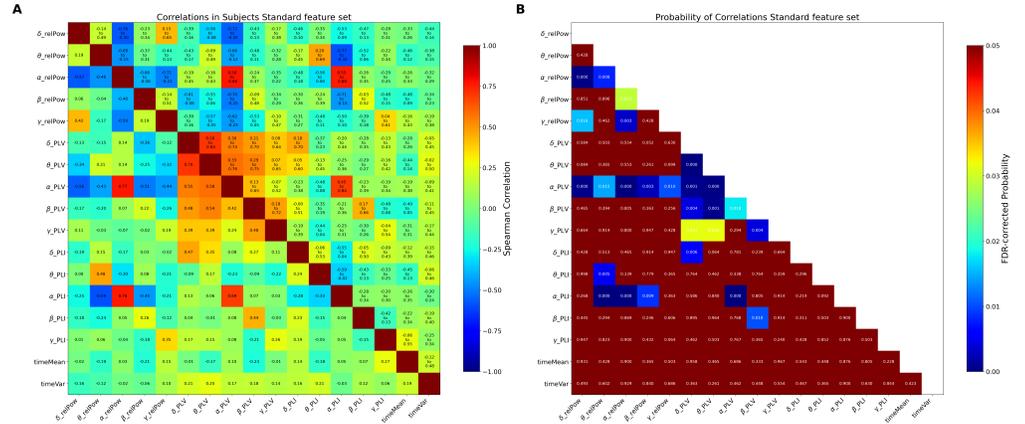}
	\captionsetup{justification=justified}
	\caption{ Pairwise correlations among features of the Standard EEG feature set. (A) Spearman correlation matrix computed across all subjects. The lower triangle displays correlation coefficients ($\rho$), while the upper triangle shows the corresponding 95\% bootstrap confidence intervals (5000 resamples). (B) Matrix of FDR-corrected p-values (Benjamini--Hochberg) for the same pairwise comparisons, shown for the lower triangle only. The Standard feature set includes relative spectral power ($\delta$, $\theta$, $\alpha$, $\beta$, $\gamma$ bands), phase synchronization metrics—phase locking value (PLV) and phase lag index (PLI) for the same frequency bands—and time-domain signal statistics (mean voltage and variance). Relative power represents the percentage of each frequency band power to the total spectral power.}
\end{figure}

\begin{figure}[h!]
	\centering
	\includegraphics[width=1.0\textwidth]{Dougalis_2026_Fig5.png}
	\captionsetup{justification=justified}
	\caption{ Pairwise correlations among features of the Dynamical EEG feature set. (A) Spearman correlation matrix computed across all subjects. The lower triangle displays correlation coefficients ($\rho$), while the upper triangle shows the corresponding 95\% bootstrap confidence intervals (5000 resamples). (B) Matrix of FDR-corrected p-values (Benjamini--Hochberg) for the same pairwise comparisons, shown for the lower triangle only. The Dynamical feature set includes measures of spectral aperiodic activity (aperiodic exponent and offset), aperiodic-corrected spectral peak characteristics (central frequency, bandwidth, and power), long-range temporal correlations via detrended fluctuation analysis (DFA), functional excitation--inhibition balance (fEI), bicoherence phase--amplitude coupling (bicPAC), instantaneous frequency dynamics (range[rng\_freqSld], rate of change[dif\_freqSld], inter-electrode correlation[corr\_freqSld], and modulation frequency[mod\_freqSld]), harmonic locking between frequency bands (HarmLock), and neuronal avalanche statistics, including avalanche size and deviation from a power-law scaling ($\kappa$ size).}
\end{figure}

\subsection*{Group-level EEG feature differences across conditions}

Group-level comparisons revealed systematic differences between healthy controls (CN) and Parkinson’s disease patients, as well as between PD patients on and off medication. These comparisons were performed independently of the deep learning classification analyses and are reported to provide physiological context for the extracted EEG features. Analyses were performed on subject-level means across all electrodes. Pairwise comparisons were controlled using FDR correction within feature families, and effect sizes (Cliff’s delta) with 95\% confidence intervals are reported in Tables 4 (Standard features) and 5 (Dynamical features).

\begin{table}[htbp]
	\centering
	\scriptsize
	\renewcommand{\arraystretch}{1.2}
	
	\resizebox{\textwidth}{!}{%
		\begin{tabularx}{\textwidth}{*{4}{>{\centering\arraybackslash}X}}
			\toprule
			
			\multicolumn{4}{c}{\textbf{Cliff’s delta or rank-biserial corr, CI [low, high], p-val FDR-corrected}} \\
			
			\midrule
			\textbf{Standard Features} & \textbf{CN vs PDoff} & \textbf{CN vs PDon} & \textbf{PDoff vs PDon} \\
			
			\midrule
			Relative Delta Power 
			& \makecell[c]{0.325 [-0.46, 1.3] \\ 0.6160} 
			& \makecell[c]{0.4 [-0.32, 1.41] \\ 0.1445} 
			& \makecell[c]{0.2 [-0.14, 0.33] \\ 0.3461} \\
			
			Relative Theta Power 
			& \makecell[c]{-0.225 [-1.04, 0.26] \\ 0.7147} 
			& \makecell[c]{-0.458 [-1.43, -0.17] \\ 0.1445} 
			& \makecell[c]{-0.467 [-0.84, -0.13] \\ 0.0923} \\
			
			Relative Alpha Power
			& \makecell[c]{0.025 [-0.83, 0.92] \\ 0.9056} 
			& \makecell[c]{0.0 [-0.79, 0.95] \\ 1.0000} 
			& \makecell[c]{0.067 [-0.2, 0.26] \\ 0.8469} \\
			
			Relative Beta Power
			& \makecell[c]{0.1 [-0.45, 1.03] \\ 0.7941} 
			& \makecell[c]{0.0 [-0.89, 0.8] \\ 1.0000} 
			& \makecell[c]{-0.467 [-0.74, -0.02] \\ 0.0923} \\
			
			Relative Gamma Power
			& \makecell[c]{-0.125 [-1.0, 0.69] \\ 0.7941} 
			& \makecell[c]{0.242 [-0.46, 1.16] \\ 0.4195} 
			& \makecell[c]{0.467 [0.13, 1.04] \\ 0.0923} \\
			
			Absolute Delta Power
			& \makecell[c]{-0.275 [-1.22, 0.22] \\ 0.1921} 
			& \makecell[c]{0.108 [-0.37, 0.96] \\ 0.7591} 
			& \makecell[c]{0.867 [0.47, 1.23] \\ 0.0005} \\
			
			Absolute Theta Power
			& \makecell[c]{-0.492 [-1.39, -0.42] \\ 0.0672} 
			& \makecell[c]{-0.492 [-1.33, -0.28] \\ 0.0985} 
			& \makecell[c]{0.067 [-0.13, 0.35] \\ 0.7615} \\
			
			Absolute Alpha Power
			& \makecell[c]{-0.358 [-1.56, 0.28] \\ 0.1115} 
			& \makecell[c]{-0.258 [-1.49, 0.53] \\ 0.4765} 
			& \makecell[c]{0.733 [0.06, 0.54] \\ 0.0168} \\
			
			Absolute Beta Power
			& \makecell[c]{-0.383 [-1.58, -0.01] \\ 0.1115} 
			& \makecell[c]{-0.225 [-1.23, 0.27] \\ 0.4765} 
			& \makecell[c]{0.467 [0.05, 0.68] \\ 0.0441} \\
			
			Absolute Gamma Power
			& \makecell[c]{-0.467 [-1.37, -0.15] \\ 0.0672} 
			& \makecell[c]{0.017 [-0.92, 0.56] \\ 0.9370} 
			& \makecell[c]{0.6 [0.1, 1.04] \\ 0.0441} \\
			
			Delta PLV
			& \makecell[c]{-0.525 [-1.68, -0.43] \\ 0.0295} 
			& \makecell[c]{-0.367 [-1.28, 0.13] \\ 0.1025} 
			& \makecell[c]{0.333 [-0.33, 0.95] \\ 0.8904} \\
			
			Theta PLV
			& \makecell[c]{-0.675 [-2.6, -0.66] \\ 0.0070} 
			& \makecell[c]{-0.667 [-2.32, -0.74] \\ 0.0080} 
			& \makecell[c]{-0.067 [-0.28, 0.56] \\ 0.8904} \\
			
			Alpha PLV
			& \makecell[c]{-0.25 [-1.21, 0.47] \\ 0.2357} 
			& \makecell[c]{-0.292 [-1.24, 0.48] \\ 0.1665} 
			& \makecell[c]{-0.2 [-0.25, 0.35] \\ 0.8904} \\
			
			Beta PLV
			& \makecell[c]{-0.342 [-1.53, 0.16] \\ 0.1314} 
			& \makecell[c]{-0.408 [-1.81, -0.06] \\ 0.0880} 
			& \makecell[c]{-0.467 [-0.84, 0.0] \\ 0.4165} \\
			
			Gamma PLV
			& \makecell[c]{-0.5 [-1.71, 0.06] \\ 0.0295} 
			& \makecell[c]{-0.458 [-1.74, 0.17] \\ 0.0742} 
			& \makecell[c]{0.067 [-0.35, 0.61] \\ 0.8904} \\
			
			Delta PLI
			& \makecell[c]{-0.217 [-1.01, -0.0] \\ 0.5713} 
			& \makecell[c]{-0.325 [-0.99, -0.23] \\ 0.6160} 
			& \makecell[c]{-0.067 [-0.94, 0.25] \\ 0.9780} \\
			
			Theta PLI
			& \makecell[c]{-0.2 [-0.87, 0.22] \\ 0.5713} 
			& \makecell[c]{0.025 [-0.75, 0.54] \\ 0.9056} 
			& \makecell[c]{-0.067 [-0.28, 0.57] \\ 0.9780} \\
			
			Alpha PLI
			& \makecell[c]{0.133 [-0.5, 1.17] \\ 0.5800} 
			& \makecell[c]{0.208 [-0.24, 1.28] \\ 0.6408} 
			& \makecell[c]{0.2 [-0.01, 0.46] \\ 0.4220} \\
			
			Beta PLI
			& \makecell[c]{0.3 [-0.05, 1.22] \\ 0.5713} 
			& \makecell[c]{-0.108 [-0.71, 0.57] \\ 0.7591} 
			& \makecell[c]{-0.733 [-1.46, -0.38] \\ 0.0100} \\
			
			Gamma PLI
			& \makecell[c]{-0.117 [-1.13, 0.64] \\ 0.5800} 
			& \makecell[c]{-0.183 [-1.18, 0.57] \\ 0.6408} 
			& \makecell[c]{-0.2 [-0.9, 0.58] \\ 0.9780} \\
			
			Time-Domain Voltage Mean
			& \makecell[c]{0.3 [-0.09, 1.36] \\ 0.1547} 
			& \makecell[c]{0.125 [-0.47, 0.89] \\ 0.5532} 
			& \makecell[c]{-0.333 [-1.14, 0.08] \\ 0.3028} \\
			
			Time-Domain Voltage Variance
			& \makecell[c]{-0.483 [-1.6, -0.29] \\ 0.0329} 
			& \makecell[c]{-0.192 [-1.03, 0.41] \\ 0.5450} 
			& \makecell[c]{0.733 [0.26, 1.08] \\ 0.0051} \\
			
			\bottomrule
		\end{tabularx}
	}
	
	\caption{Comparative summary of group-level comparisons using the Standard feature set across diagnostic contrasts (unpaired: CN–PDoff, CN–PDon; paired: PDoff–PDon). Values represent Cliff’s delta or matched-pairs rank-biserial correlation (unpaired or paired respectively) with 95\% confidence intervals and Wilcoxon (rank sum or signed rank sum) FDR-corrected p-values (Benjamini–Hochberg method).}
	
\end{table}

\begin{table}[htbp]
	\centering
	\scriptsize
	\renewcommand{\arraystretch}{1.2}
	
	\resizebox{\textwidth}{!}{%
		\begin{tabularx}{\textwidth}{*{4}{>{\centering\arraybackslash}X}}
			\toprule
			\multicolumn{4}{c}{\textbf{Cliff’s delta or rank-biserial corr, CI [low, high], p-val FDR-corrected}} \\
			\midrule
			\textbf{Dynamical Features} & \textbf{CN vs PDoff} & \textbf{CN vs PDon} & \textbf{PDoff vs PDon} \\
			\midrule
			Aperiodic exponent
			& \makecell[c]{0.3 [-0.12, 1.33] \\ 0.4561} 
			& \makecell[c]{0.092 [-0.41, 0.9] \\ 0.6637} 
			& \makecell[c]{-0.333 [-0.9, 0.59] \\ 0.3894} \\
			Aperiodic offset
			& \makecell[c]{-0.217 [-1.27, 0.27] \\ 0.4561} 
			& \makecell[c]{-0.1 [-0.7, 0.72] \\ 0.6637} 
			& \makecell[c]{0.067 [0.07, 0.81] \\ 0.3786} \\
			Periodic Peak Central Frequency (CF)
			& \makecell[c]{0.308 [-0.06, 1.41] \\ 0.3633} 
			& \makecell[c]{0.483 [0.33, 1.65] \\ 0.0657} 
			& \makecell[c]{0.2 [-0.08, 1.0] \\ 0.5995} \\
			Periodic Peak Band power (BP)
			& \makecell[c]{-0.192 [-1.3, 0.31] \\ 0.3633} 
			& \makecell[c]{-0.275 [-1.63, 0.24] \\ 0.2059} 
			& \makecell[c]{-0.2 [-0.59, 0.11] \\ 0.5995} \\
			Periodic Peak Bandwidth (BW)
			& \makecell[c]{0.192 [-0.58, 1.17] \\ 0.3633} 
			& \makecell[c]{0.267 [-0.37, 1.44] \\ 0.2059} 
			& \makecell[c]{0.067 [-0.16, 0.59] \\ 0.5995} \\
			Delta DFA exponent
			& \makecell[c]{-0.067 [-0.8, 0.46] \\ 0.7820} 
			& \makecell[c]{-0.05 [-0.85, 0.63] \\ 0.9056} 
			& \makecell[c]{0.067 [-0.68, 0.77] \\ 0.9341} \\
			Theta DFA exponent
			& \makecell[c]{-0.35 [-1.56, 0.15] \\ 0.2422} 
			& \makecell[c]{-0.333 [-1.58, 0.13] \\ 0.2053} 
			& \makecell[c]{-0.2 [-0.56, 0.43] \\ 0.9341} \\
			Alpha DFA exponent
			& \makecell[c]{-0.058 [-0.82, 0.86] \\ 0.7820} 
			& \makecell[c]{0.025 [-0.89, 0.87] \\ 0.9056} 
			& \makecell[c]{0.067 [-0.52, 0.45] \\ 0.9341} \\
			Beta DFA exponent
			& \makecell[c]{-0.283 [-1.27, 0.23] \\ 0.2983} 
			& \makecell[c]{-0.325 [-1.44, 0.09] \\ 0.2053} 
			& \makecell[c]{-0.333 [-0.82, 0.51] \\ 0.9341} \\
			Gamma DFA exponent
			& \makecell[c]{-0.408 [-1.5, -0.04] \\ 0.2422} 
			& \makecell[c]{-0.333 [-1.5, 0.22] \\ 0.2053} 
			& \makecell[c]{0.067 [-0.41, 0.72] \\ 0.9341} \\
			Delta fE/I
			& \makecell[c]{-0.108 [-0.95, 0.58] \\ 0.7518} 
			& \makecell[c]{-0.067 [-0.98, 0.49] \\ 0.7518} 
			& \makecell[c]{-0.2 [-0.76, 0.58] \\ 0.9780} \\
			Theta fE/I
			& \makecell[c]{-0.15 [-1.09, 0.38] \\ 0.7518} 
			& \makecell[c]{-0.275 [-1.41, 0.22] \\ 0.6710} 
			& \makecell[c]{0.067 [-0.78, 0.36] \\ 0.9780} \\
			Alpha fE/I
			& \makecell[c]{-0.167 [-1.1, 0.39] \\ 0.7518} 
			& \makecell[c]{-0.233 [-1.44, 0.51] \\ 0.6710} 
			& \makecell[c]{-0.067 [-0.79, 0.55] \\ 0.9780} \\
			Beta fE/I
			& \makecell[c]{0.067 [-0.51, 0.9] \\ 0.7518} 
			& \makecell[c]{0.15 [-0.47, 1.13] \\ 0.7518} 
			& \makecell[c]{0.333 [-0.53, 0.84] \\ 0.9780} \\
			Gamma fE/I
			& \makecell[c]{0.175 [-0.51, 1.24] \\ 0.7518} 
			& \makecell[c]{-0.067 [-0.57, 0.91] \\ 0.7518} 
			& \makecell[c]{-0.333 [-0.62, 0.32] \\ 0.9780} \\
			Delta/gamma PAC
			& \makecell[c]{-0.192 [-1.16, 0.55] \\ 0.7907} 
			& \makecell[c]{0.075 [-0.85, 0.93] \\ 0.9370} 
			& \makecell[c]{-0.067 [-0.6, 0.85] \\ 0.9341} \\
			Theta/gamma PAC
			& \makecell[c]{0.008 [-0.91, 0.59] \\ 0.9685} 
			& \makecell[c]{-0.017 [-0.87, 0.47] \\ 0.9370} 
			& \makecell[c]{-0.067 [-0.64, 0.42] \\ 0.9138} \\
			Theta IF range
			& \makecell[c]{0.042 [-0.48, 0.75] \\ 0.8433} 
			& \makecell[c]{0.442 [0.17, 1.33] \\ 0.1086} 
			& \makecell[c]{0.733 [0.13, 1.03] \\ 0.0045} \\
			Alpha IF range
			& \makecell[c]{-0.142 [-1.1, 0.67] \\ 0.7524} 
			& \makecell[c]{-0.008 [-0.96, 0.65] \\ 0.9685} 
			& \makecell[c]{0.333 [-0.25, 0.4] \\ 0.5995} \\
			Gamma IF range
			& \makecell[c]{-0.3 [-1.21, 0.17] \\ 0.4641} 
			& \makecell[c]{-0.15 [-1.03, 0.6] \\ 0.7152} 
			& \makecell[c]{0.067 [-0.26, 0.88] \\ 0.5995} \\
			Theta IF rate of change
			& \makecell[c]{0.517 [0.25, 1.31] \\ 0.0429} 
			& \makecell[c]{0.583 [0.48, 1.77] \\ 0.0171} 
			& \makecell[c]{0.333 [0.04, 0.72] \\ 0.1662} \\
			Alpha IF rate of change
			& \makecell[c]{-0.117 [-0.96, 0.75] \\ 0.5800} 
			& \makecell[c]{-0.025 [-0.91, 0.79] \\ 0.9056} 
			& \makecell[c]{-0.067 [-0.25, 0.33] \\ 0.8469} \\
			Gamma IF rate of change
			& \makecell[c]{-0.142 [-0.84, 0.54] \\ 0.5800} 
			& \makecell[c]{0.083 [-0.68, 0.72] \\ 0.9056} 
			& \makecell[c]{0.333 [-0.12, 0.8] \\ 0.2814} \\
			Theta IF frequency
			& \makecell[c]{-0.3 [-1.46, 0.24] \\ 0.2498} 
			& \makecell[c]{-0.108 [-0.85, 0.43] \\ 0.9056} 
			& \makecell[c]{0.2 [-0.16, 0.99] \\ 0.6387} \\
			Alpha IF frequency
			& \makecell[c]{-0.292 [-1.19, 0.05] \\ 0.2498} 
			& \makecell[c]{-0.025 [-0.93, 0.7] \\ 0.9056} 
			& \makecell[c]{0.067 [-0.44, 1.12] \\ 0.6387} \\
			Gamma IF frequency
			& \makecell[c]{0.1 [-0.62, 1.2] \\ 0.6353} 
			& \makecell[c]{0.292 [-0.35, 1.36] \\ 0.4995} 
			& \makecell[c]{0.2 [-0.56, 0.98] \\ 0.6387} \\
			Alpha/theta IF harmonic lock
			& \makecell[c]{0.558 [0.44, 2.16] \\ 0.0162} 
			& \makecell[c]{0.575 [0.47, 2.15] \\ 0.0108} 
			& \makecell[c]{0.067 [-0.09, 0.12] \\ 0.7615} \\
			Gamma/theta IF harmonic lock
			& \makecell[c]{-0.5 [-2.0, -0.21] \\ 0.0236} 
			& \makecell[c]{-0.558 [-2.19, -0.41] \\ 0.0108} 
			& \makecell[c]{0.067 [-0.26, 0.19] \\ 0.7615} \\
			Avalanche Duration
			& \makecell[c]{0.308 [-0.44, 1.15] \\ 0.2321} 
			& \makecell[c]{-0.167 [-1.14, 0.25] \\ 0.4526} 
			& \makecell[c]{-0.6 [-1.46, -0.26] \\ 0.0188} \\
			Kappa Size
			& \makecell[c]{0.183 [-0.38, 0.93] \\ 0.3845} 
			& \makecell[c]{-0.158 [-1.19, 0.32] \\ 0.4526} 
			& \makecell[c]{-0.733 [-1.22, -0.39] \\ 0.0045} \\
			\bottomrule
		\end{tabularx}%
	} 
	
	\caption{Comparative summary of group-level comparisons using the Dynamical feature set across diagnostic contrasts (unpaired: CN–PDoff, CN–PDon; paired: PDoff–PDon). Values represent Cliff’s delta or matched-pairs rank-biserial correlation (unpaired or paired respectively) with 95\% confidence intervals and Wilcoxon (rank sum or signed rank sum) FDR-corrected p-values (Benjamini–Hochberg method).}
\end{table}

Across the Standard feature set, comparison between PD\textsubscript{off} vs PD\textsubscript{on} contrast captured medication-related effects. Specifically, absolute delta power and voltage variance were higher in PD\textsubscript{off} compared to PD\textsubscript{on} groups, indicating that dopaminergic therapy reduced slow-wave power and overall voltage variability (Figure 6A, B). Similarly, other spectral bands (alpha, beta and gamma) exhibited modest medication-induced decreases, suggesting normalization of high-frequency activity with therapy. Notably, relative spectral power did not reproduce the medication-related differences observed in absolute spectral power. In the PD\textsubscript{off} vs PD\textsubscript{on} contrast, relative-band effects were attenuated across frequencies and did not survive correction, consistent with the normalization inherent in relative power measures that redistribute variance across bands. Beta and gamma relative power showed near-threshold effects (FDR-corrected $p \approx 0.09$), with the polarity of beta differing from the absolute metric, likely reflecting this normalization rather than a reversal of the underlying spectral change. In comparisons between CN and PD patients, prominent differences emerged primarily in theta phase synchrony (PLV), which was higher in both PD\textsubscript{off} and PD\textsubscript{on} relative to CN. Increased synchrony was also observed in the delta and gamma PLV between CN and PD patients off medication (Figure 6C). While the CN vs PD\textsubscript{on} contrasts showed trends toward lower synchrony, these differences were only marginally insignificant, indicating that the elevated synchrony largely persists despite dopaminergic therapy. Overall, the Standard feature set captured both medication-sensitive and disease-related spectral and synchrony dynamics, particularly in delta and theta bands. To evaluate whether these phase synchrony differences could reflect volume conduction rather than true neural coupling, PLV and PLI were recomputed after applying a surface Laplacian transform. PLV effects persisted following this spatial filtering, while PLI remained nonsignificant (Supplementary Table 1), indicating that the observed near-zero-lag synchrony likely reflects genuine neural interactions rather than artefactual volume conduction.

In the Dynamical features, the PD\textsubscript{off} vs PD\textsubscript{on} contrast revealed medication-related modulation of theta-band dynamics. The effective frequency range of the theta instantaneous frequency (IF), quantified through frequency sliding, decreased with medication, while the critical dynamics quantified via neuronal avalanche duration and deviation from power law distribution (kappa size) metrics were elevated suggesting a recovery of theta spectro-temporal stability and increase in the large-scale network criticality (Figure 6D, F).  The rate of change of theta IF was lower in PD patients (both on or off medication) compared to controls, indicating decreased responsiveness of temporal oscillatory variability in pathology that is resistant to medication. Measures of cross-frequency coupling displayed distinct disease-specific patterns: alpha/theta harmonic locking was reduced in PD patients relative to controls, whereas gamma/theta harmonic locking increased, indicating a shift in large-scale cross-frequency coordination in Parkinson’s disease (Figure 6E).  Other dynamical metrics, including aperiodic components, DFA exponent or inhibition-excitation balance metrics (fE/I), did not show significant medication-related or disease-related changes at the global electrode level after FDR corrections suggesting that these properties may either be less sensitive to dopaminergic modulation or more specific to spatially restricted brain topologies.

\begin{figure}[h!]
	\centering
	\includegraphics[width=1.0\textwidth]{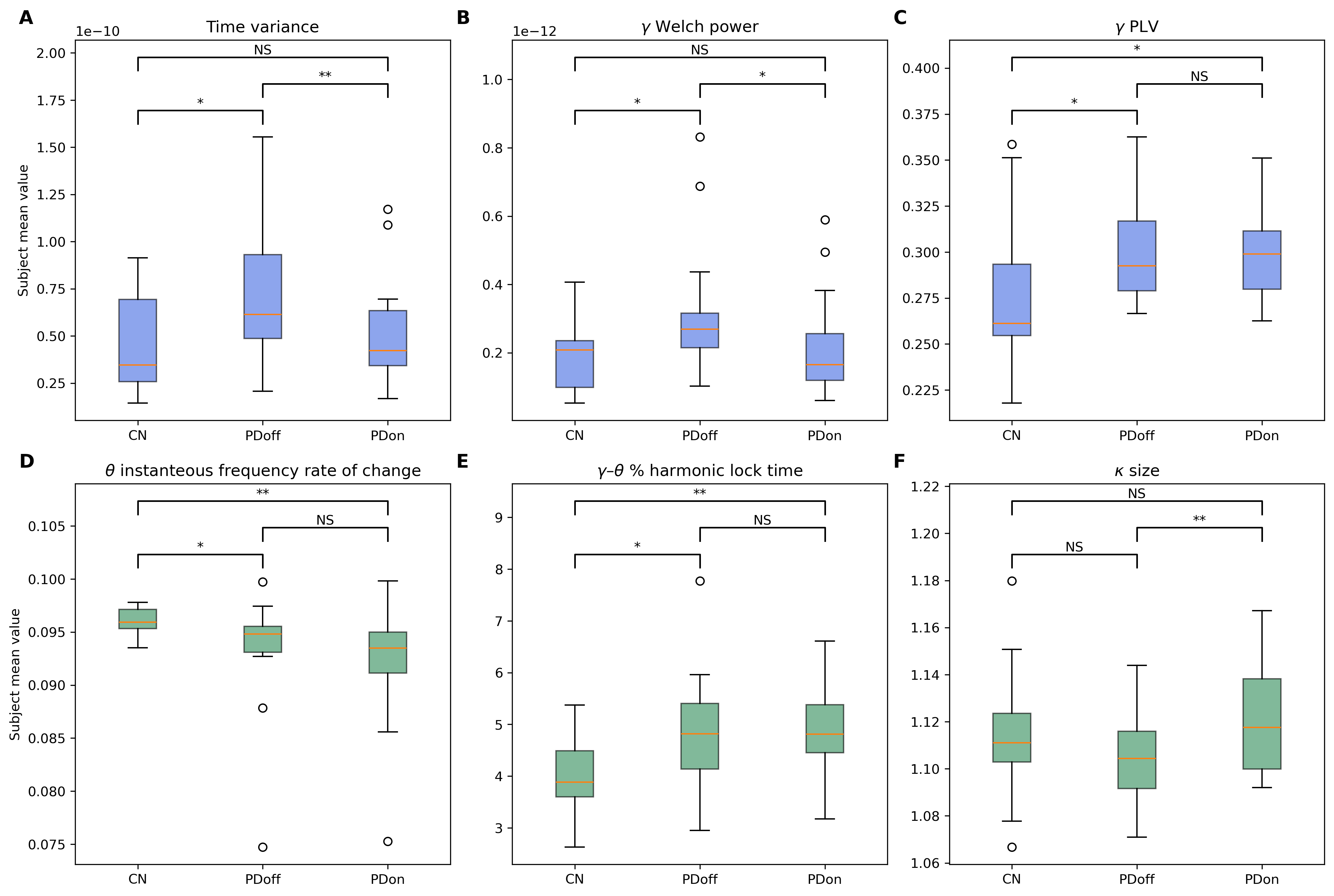}
	\captionsetup{justification=justified}
	\caption{ Group differences in representative Standard and Dynamical EEG features. Boxplots showing the distribution of selected EEG-derived features for healthy controls (CN), Parkinson’s disease patients assessed off medication (PD\textsubscript{off}), and the same patients assessed on dopaminergic medication (PD\textsubscript{on}). Panels from the Standard feature set include time-domain variance (A), absolute $\gamma$-band Welch power (B), and $\gamma$-band phase-locking value (PLV) (C). Panels from the Dynamical feature set include frequency sliding instantaneous $\theta$-band frequency rate of change (D), percentage time of $\gamma$--$\theta$ harmonic lock (E), and avalanche $\kappa$ size metric (F). In each boxplot, the central line indicates the median, the box represents the interquartile range (25th--75th percentiles), and whiskers extend to 1.5$\times$ the interquartile range. Pairwise statistical comparisons were performed using Wilcoxon rank-sum tests for independent group comparisons (CN vs PD\textsubscript{off} and CN vs PD\textsubscript{on}) and Wilcoxon signed-rank tests for the paired comparison between medication states (PD\textsubscript{off} vs PD\textsubscript{on}). Multiple comparisons were controlled using false discovery rate (FDR) correction within feature families. Significant comparisons are indicated by asterisks (NS, not significant $>$ 0.05, *p $<$ 0.05, **p $<$ 0.01).}
\end{figure}

Taken together, these results indicate complementary contributions of Standard and Dynamical features in the delineation of differences of PD states. Standard features were particularly informative in capturing medication-related effects (e.g., delta power, voltage variance), while Dynamical features highlighted disease-specific network alterations (e.g., harmonic locking, Kappa size, theta IF dynamics). Across all contrasts, on a global electrode analysis, the directionality of Cliff’s delta supports the interpretation that dopaminergic therapy dampens to a different degree both slow-wave \& higher frequency spectral activity whereas PD is associated with enhanced multiband synchrony and altered cross-frequency interactions relative to healthy controls, features that are unresponsive to correction by dopaminergic medication.

\section*{Discussion}

The present study examined whether physiologically interpretable EEG features capturing complementary aspects of neural dynamics can discriminate Parkinsonian neural states under a strict subject-level validation framework. In addition to evaluating classification performance using transformer-based models, the study also examined group-level differences in these electrophysiological descriptors to provide a physiological interpretation of the EEG features that contribute to the discrimination between Parkinson’s disease patients and healthy controls and between medication states. Using a transformer-based classifier trained on two conceptually distinct feature families—Standard electrophysiological descriptors and Dynamical measures of neural signal organization—classification performance was evaluated across multiple diagnostic contrasts and examined the informational structure of the extracted feature sets. The results indicate that multivariate EEG representations contain discriminative information about Parkinsonian neural activity, while also revealing that different feature families contribute preferentially to different diagnostic contrasts.

\subsection*{Multivariate EEG representations capture discriminative information about Parkinsonian neural states}

Consistent with the central hypothesis of this work, EEG-derived feature representations enabled classification between Parkinson’s disease patients and cognitively normal individuals, as well as between medication states within the same patients. Although classification performance varied across contrasts, the results demonstrate that distributed electrophysiological descriptors extracted from resting-state EEG contain measurable information related to Parkinsonian neural dynamics as seen previously by others (Bunterngchit et al., 2025; Mukherjee \& Roy, 2025). Importantly, these results support the notion that EEG signatures of Parkinson’s disease are unlikely to manifest as single robust biomarkers, but instead may emerge from the joint contribution of multiple partially independent signal features. This observation is consistent with prior reports suggesting that electrophysiological alterations in Parkinson’s disease are spatially heterogeneous, spectrally distributed, and dependent on the specific analytic metric employed (Babiloni et al., 2011; Jackson et al., 2019; Waninger et al., 2020). By integrating a broad range of interpretable descriptors—including spectral, connectivity, and dynamical signal properties—the present study adopts a multivariate perspective that may better capture the complex alterations in cortical activity associated with basal ganglia dysfunction.

\subsection*{Differential contribution of Standard and Dynamical EEG descriptors}

A central aim of the study was to examine whether traditional EEG descriptors and dynamical measures of neural activity provide complementary information about Parkinsonian brain states. The results suggest that the relative contribution of these feature families depends on the specific classification problem. For the discrimination between PD patients off and on dopaminergic medication, the Standard feature configuration—comprising spectral power and synchronization measures—produced the strongest classification performance and significantly outperformed the Dynamical feature set. This observation may reflect the well-established sensitivity of oscillatory synchronization and spectral power changes to dopaminergic modulation within basal ganglia–thalamo–cortical circuits. Dopamine replacement therapy and/or deep brain stimulations are known to alter functional brain gradients (Orlando et al., 2025), cortical beta-band oscillatory activity and network synchronization (Silberstein et al., 2005; George et al., 2013; Miller et al., 2019) or basal ganglia-thalamus-cortical network activity (de Hemptinne et al., 2015), phenomena that are often detectable through conventional spectral and connectivity metrics (Stoffers et al., 2007; Jackson et al., 2019; Zhang et al., 2022).

In contrast, the Dynamical feature set performed competitively with the Standard configuration in contrasts involving healthy controls and Parkinson’s disease patients. These features quantify higher-order aspects of neural activity such as scale-free temporal correlations, cross-frequency interactions, and neuronal avalanche statistics. Such descriptors may capture broader changes in the organization of neural population activity rather than specific oscillatory signatures associated with dopaminergic tone. This interpretation is consistent with theoretical perspectives suggesting that neurodegenerative processes can alter the dynamical regime of neural networks (Kim et al., 2017; Sadeghi et al., 2025), including properties related to criticality (Zimmern, 2020), excitation–inhibition balance (Mohanty et al., 2025), and large-scale coordination of activity (Zhu et al., 2021; Akgüller et al., 2024).

Importantly, these findings do not indicate that Dynamical features are inferior to traditional EEG metrics. Rather, the results suggest that different feature families capture distinct physiological aspects of Parkinsonian neural activity, with Standard descriptors appearing particularly sensitive to medication-induced changes in oscillatory synchronization, while Dynamical descriptors capture broader alterations in neural signal organization.

\subsection*{Evidence for complementary information within dynamical feature representations}

The feature ablation analysis provides additional insight into the informational structure of the two feature configurations. In contrasts involving healthy controls and Parkinson’s disease patients, the Dynamical feature configuration showed a consistent pattern in which the full model accuracy exceeded the mean accuracy obtained from randomly reduced feature subsets. This indicates that the predictive performance of the Dynamical configuration depends on the combined contribution of multiple features rather than a small number of dominant descriptors. In contrast, the Standard feature configuration showed comparatively weaker separation between the full model and randomly reduced feature subsets in these contrasts. This suggests that removing random subsets of traditional spectral and synchronization features does not substantially degrade classification performance, implying that these descriptors may contain partially overlapping information. Taken together, these results suggest that dynamical descriptors capture complementary aspects of neural signal organization, such that their joint inclusion improves classification performance. This observation aligns with theoretical views of brain activity as a complex dynamical system in which multiple interacting processes—oscillatory, scale-free, and network-level dynamics—collectively shape observed electrophysiological signals (Hardstone et al., 2012; Zimmern, 2020).

\subsection*{Low redundancy among extracted EEG features}

The correlation analysis further supports the interpretation that the extracted features represent largely distinct descriptors of EEG dynamics. Across both feature configurations, most pairwise correlations were moderate, with only a small fraction of feature pairs exhibiting strong associations. Importantly, the majority of pairwise correlations remained well below levels typically considered indicative of strong redundancy.  These observations confirm that the feature sets were constructed to capture complementary rather than redundant information about neural activity. Even within related methodological families—such as spectral power and phase synchronization—correlations remained moderate rather than approaching unity. The Dynamical feature set similarly showed limited redundancy despite including multiple descriptors derived from related analytical frameworks. Together with the ablation results, this finding supports the interpretation that classification performance arises from the integration of multiple partially independent electrophysiological descriptors, rather than from isolated features dominating model predictions.

\subsection*{Physiological interpretation of group-level EEG differences}

Inspection of the group-level contrasts provides additional context for the classification results by clarifying which electrophysiological properties differ between Parkinson’s disease (PD) patients and controls and which are modulated by dopaminergic therapy. Importantly, the group-level comparisons reported here were conducted independently of the feature-selection process used for classification. The EEG features included in the model were defined a priori based on physiologically motivated descriptors of neural dynamics, and the statistical comparisons were performed post hoc to aid physiological interpretation of the feature representations. Several effects were clearly sensitive to medication state. In particular, dopaminergic therapy reduced absolute delta power and global voltage variance, indicating that the unmedicated Parkinsonian state is characterized by stronger slow-wave activity and greater amplitude variability. Medication also reduced the range of theta instantaneous frequency fluctuations, suggesting stabilization of oscillatory dynamics. In contrast, neuronal avalanche statistics showed increased kappa size under medication, consistent with a shift toward larger-scale activity cascades and a more prominent supercritical dynamical regime. Together, these findings suggest that dopamine primarily modulates global spectral amplitude and aspects of large-scale network stability.

Other electrophysiological differences appeared to reflect disease-related network alterations that were largely resistant to dopaminergic normalization. Most notably, theta-band phase synchrony was elevated in PD patients both on and off medication relative to controls, indicating increased large-scale coupling in this frequency band. Additional synchrony differences were observed in the delta and gamma bands in the unmedicated state with the corresponding CN vs PD contrasts showing only marginal differences, indicating that elevated synchrony largely persists despite therapy. PLV/PLI contrasts suggested that these effects depend on near-zero-lag coupling. Laplacian filtering confirmed that PLV effects persisted while PLI did not show any differences, indicating that the observed synchrony likely reflects genuine neural coupling rather than volume conduction. Several dynamical descriptors also showed persistent disease-related alterations. The rate of change of theta instantaneous frequency was reduced in PD patients regardless of medication status, indicating slower modulation of oscillatory dynamics. In addition, cross-frequency coupling patterns differed systematically from controls, with reduced alpha--theta harmonic locking and increased gamma--theta coupling, both of which remained largely unchanged with dopaminergic therapy.

Taken together, these findings suggest a distinction between electrophysiological properties that primarily reflect dopaminergic state, such as slow-wave power, signal variability and avalanche dynamics, and those that reflect more stable disease-related changes in cortical network coordination, including altered theta synchrony and cross-frequency interactions. This pattern is consistent with the classification analyses, where traditional spectral descriptors showed the strongest sensitivity to medication contrasts, whereas dynamical descriptors performed comparably in disease classification but were less sensitive to medication effects. Within the dynamical feature set, instantaneous frequency dynamics and neuronal avalanche statistics primarily reflected medication state, while cross-frequency coupling measures differentiated Parkinson’s disease patients from healthy controls.

\subsection*{Implications for EEG biomarker development in Parkinson’s disease}

The results of the present study contribute to ongoing efforts to identify EEG-based biomarkers for Parkinson’s disease. Beyond traditional spectral and connectivity features, this study also examines dynamical descriptors of cortical activity that have not previously been systematically characterized in Parkinson’s disease using EEG. Notably, theta instantaneous frequency dynamics and neuronal avalanche statistics (duration and kappa size) differentiated medication states (PD\textsubscript{off} vs PD\textsubscript{on}), indicating sensitivity to dopaminergic modulation of large-scale cortical dynamics. In contrast, cross-frequency coupling measures involving instantaneous frequency (reduced alpha--theta harmonic locking and increased gamma--theta coupling) differentiated Parkinson’s disease patients from healthy controls, suggesting that these interactions reflect more stable disease-related alterations in network coordination. While many previous studies have focused on specific oscillatory/connectivity features or spatially localized spectral changes (Babiloni et al., 2011; Swann et al., 2015; Jackson et al., 2019; Zhang et al., 2022), the current findings suggest that global multivariate electrophysiological representations may provide a more informative characterization of Parkinsonian brain dynamics. Importantly, the use of a strict leave-one-subject-out validation framework ensures that classification results reflect generalizable subject-level patterns rather than within-subject overfitting. This methodological choice addresses a common limitation in EEG classification studies, where subject-level independence is not always maintained during model evaluation. From a translational perspective, EEG-based multivariate models could potentially complement existing clinical assessments by providing non-invasive measures sensitive to both disease-related neural alterations and medication effects. Such approaches may ultimately contribute to objective biomarkers capable of tracking disease progression or treatment response.

\subsection*{Limitations and Future Directions}

A number of limitations should be considered when interpreting the present findings. First, the analysis was conducted on a single publicly available resting-state EEG dataset, which limits the ability to assess the generalizability of the proposed feature representations across independent cohorts, recording systems, and clinical populations. Replication in larger multi-center datasets will therefore be necessary to determine the robustness of the observed electrophysiological patterns. Second, the present study focused on global electrode-level summaries of EEG features, which may obscure spatially localized cortical effects that could contribute additional discriminative information. Future work incorporating spatially resolved analyses or source-reconstructed signals may provide deeper insight into the cortical generators underlying the observed feature differences. Third, although the transformer classifier integrates multiple interpretable descriptors, the study does not attempt to infer causal mechanisms linking the extracted features to specific neurophysiological processes within basal ganglia–thalamo–cortical circuits. Finally, the dataset included only resting-state recordings, and it remains possible that task-related or movement-related EEG paradigms may reveal additional disease-specific neural signatures. Future research combining multimodal imaging, longitudinal datasets, and larger patient cohorts may help determine whether multivariate electrophysiological representations can serve as reliable biomarkers for disease progression or treatment response in Parkinson’s disease.

\subsection*{Conclusions}

In summary, this study shows that multivariate representations of resting-state EEG capture discriminative information about Parkinsonian neural states when evaluated under strict subject-level validation. Traditional spectral and synchronization descriptors were most sensitive to medication-related neural modulation, whereas dynamical descriptors—including instantaneous frequency dynamics, cross-frequency interactions, and neuronal avalanche statistics—provided complementary information reflecting both medication and disease-related network alterations. The low redundancy between these features suggests that Parkinsonian cortical dynamics are best characterized by integrating multiple interpretable electrophysiological dimensions rather than relying on single metrics. Multivariate EEG representations may therefore provide a promising non-invasive framework for studying disease-related cortical dynamics and developing objective biomarkers of Parkinson’s disease. Future validation across independent datasets will be essential to determine whether such multivariate electrophysiological representations can serve as robust clinical biomarkers.

\subsection*{Author Contributions}

A.G.D is the sole author of this manuscript and performed the following tasks: conceptualization, methodology, software development and implementation, validation, formal analysis, investigation, resources, data curation, writing—original draft preparation, writing—review and editing, visualization, project administration, and acquisition of funding.

\subsection*{Funding}

No external public or private funding sources were involved in the completion of this work.

\subsection*{Institutional Review Board Statement}

This study utilized a publicly available resting-state EEG dataset (openNeuro, Accession Number ds002778). The original study was conducted in accordance with the Institutional Review Board of the University of California, San Diego, and the Declaration of Helsinki.

\subsection*{Informed Consent Statement}

Participants provided written informed consent in accordance with the Institutional Review Board of the University of California, San Diego. For more information on the study interested readers can consult the work of George and colleagues (George et al., 2013).

\subsection*{Data Availability Statement}

This study was completed using publicly available EEG data (openNeuro, Accession Number ds002778). The scripts implementing the Multi-Head Attention Transformer architecture and the complete deep learning training and analysis routine are openly available at the author’s GitHub repository (\url{https://github.com/antoniosdougalis/Parkinson-s-Project.git}). Scripts for feature extraction routines are described in the manuscript and are available from the author upon reasonable request.

\subsection*{Acknowledgments}

The author thanks Dr. Mike X. Cohen for his invaluable guidance and educational support on deep learning algorithms, EEG analysis and statistical methods.

\subsection*{Conflicts of Interest}

The author declares no conflicts of interest.

\nocite{*}  
\printbibliography

\clearpage
\subsection*{Supplementary Materials}
\begin{figure}[h!]
	\centering
	\includegraphics[width=1.0\textwidth]{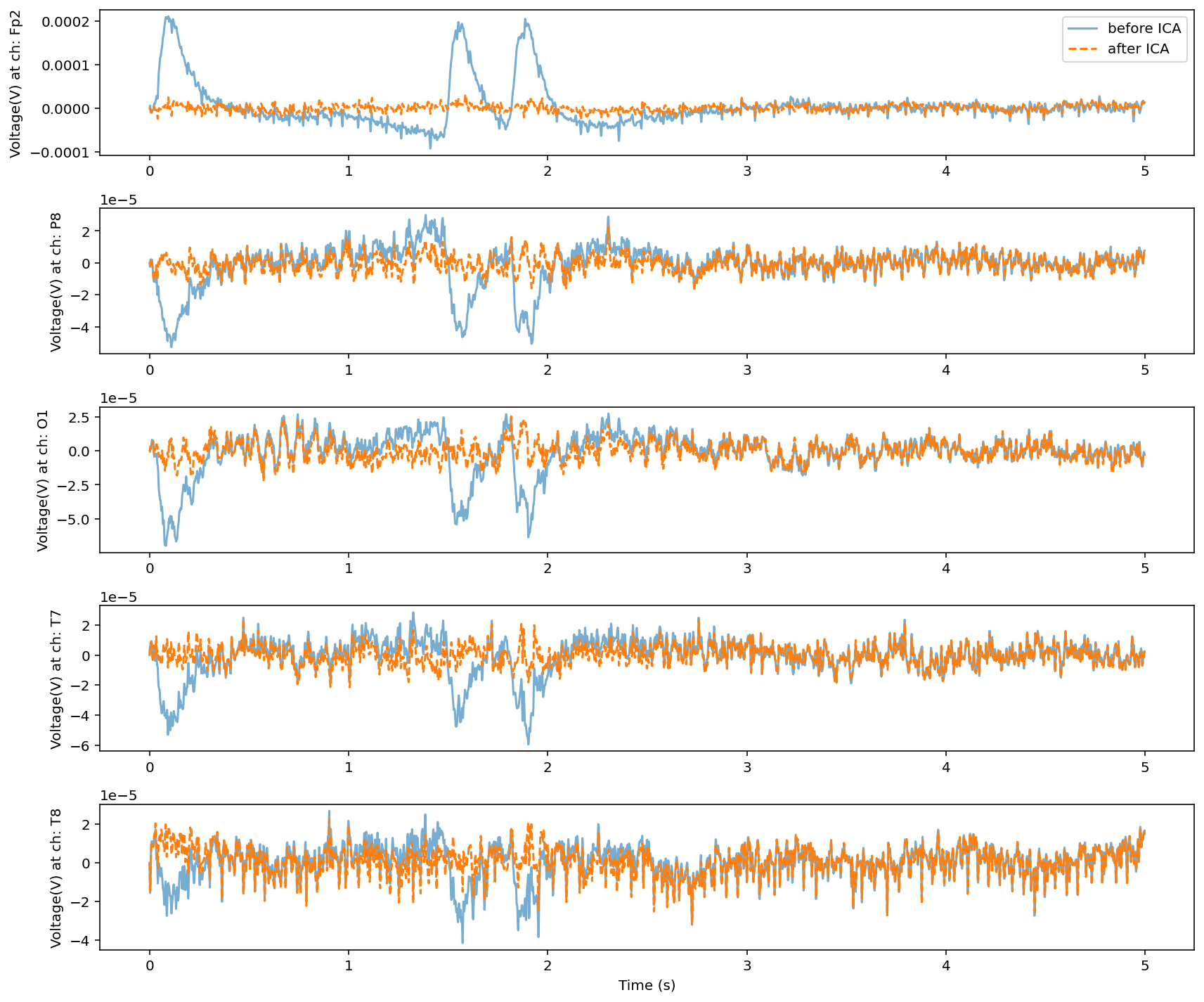}
	\captionsetup{justification=justified, labelfont=bf}
	\caption*{Supplementary Figure S1. EEG signal artifact removal via application of robust criteria and Independent Component Analysis (ICA) algorithms. Example of filtered EEG traces (1-45 Hz bandpass, blue) taken from five channels showing the resultant signal (orange) after artifact reduction via Independent Component Analysis (ICA) using a combined triple criterion based on percentile-based thresholds of the component projection power \& kurtosis and on a fixed high frequency muscle spectral power ratio (25-45Hz over 1-15Hz) to unmix neuronal and muscular sources. The component projection and kurtosis thresholds were set at the 95th percentile and the muscle spectral ratio threshold at three.}
\end{figure}

\begin{figure}[!t]
	\centering
	\includegraphics[width=1.0\textwidth]{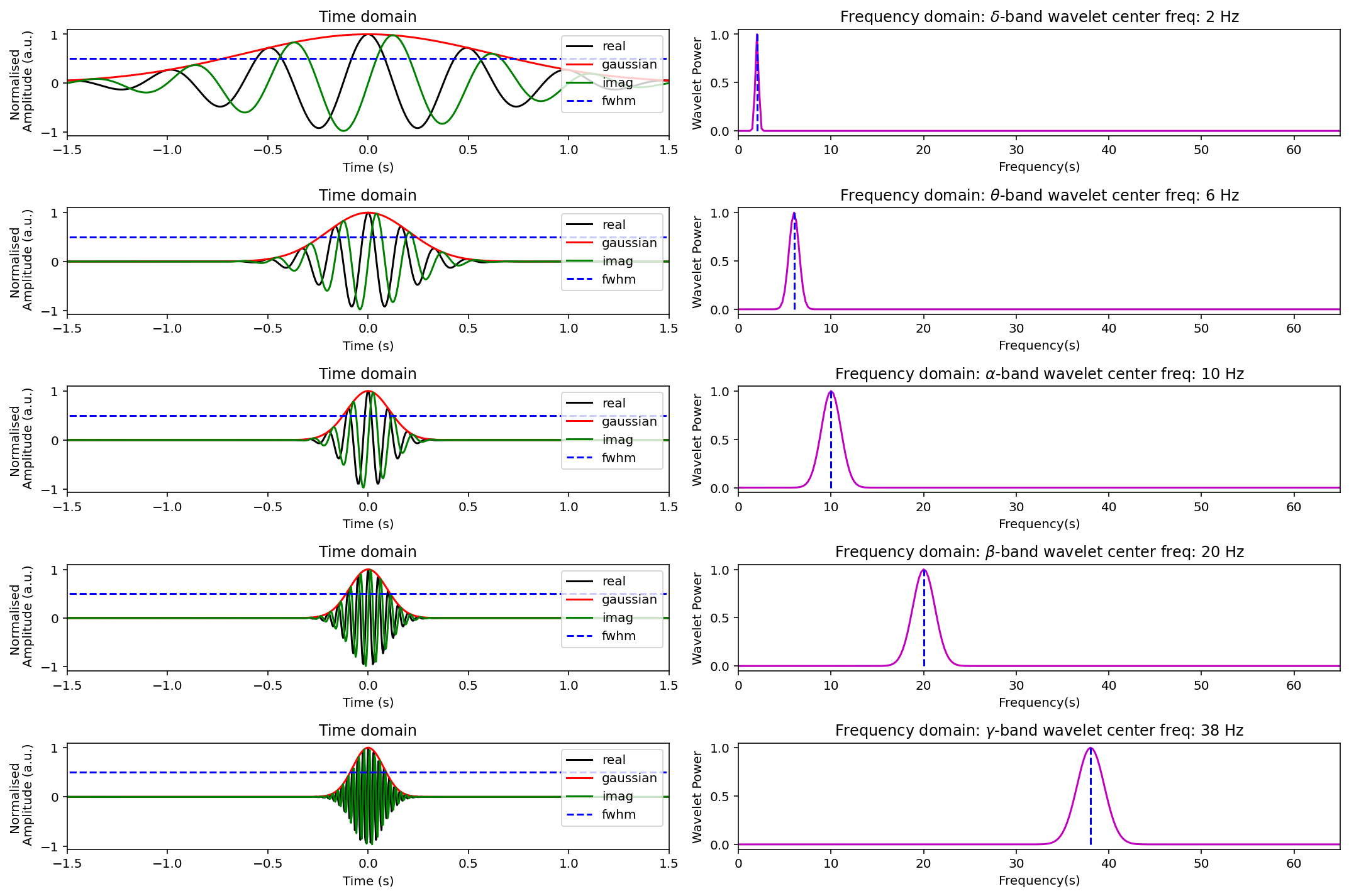}
	\captionsetup{justification=justified, labelfont=bf}
	\caption*{Supplementary Figure S2. Construction and design of complex-valued Morlet Wavelets (cMWs) for phase extraction. Complex-valued Morlet wavelets were constructed as 4-second-long complex sine waves tapered with a Gaussian window and centered at the canonical frequencies of 2, 6, 10, 20, and 38 Hz (corresponding to delta, theta, alpha, beta, and gamma bands, respectively). The Gaussian taper was defined by a fixed full width at half maximum (FWHM) in the time domain, which determines the effective bandwidth of each wavelet in the frequency domain. The left column shows the time-domain representation of each cMW, including the real and imaginary components of the complex wavelet, along with the Gaussian envelope used for tapering. The blue horizontal dashed line indicates the half-maximum level for visual estimation of the FWHM (delta, 1.45 s; theta, 0.48 s; alpha, 0.25 s; beta, 0.22 s; gamma, 0.18 s). The right column shows the corresponding frequency-domain representations, illustrating the spectral profile and effective bandwidth of each wavelet (bandwidth at half-max along central frequency[from-to range], delta, 1 Hz [from 1.50 to 2.50 Hz]; theta, 2 Hz[from 5.00 to 7.00 Hz]; alpha, 4 Hz[from 8.00 to 12.00 Hz]; beta, 4.5 Hz[from 17.75 to 22.25 Hz]; gamma, 5 Hz[from 35.50 to 40.50 Hz]).}
\end{figure}

\begin{figure}[!t]
	\centering
	\includegraphics[width=1.0\textwidth]{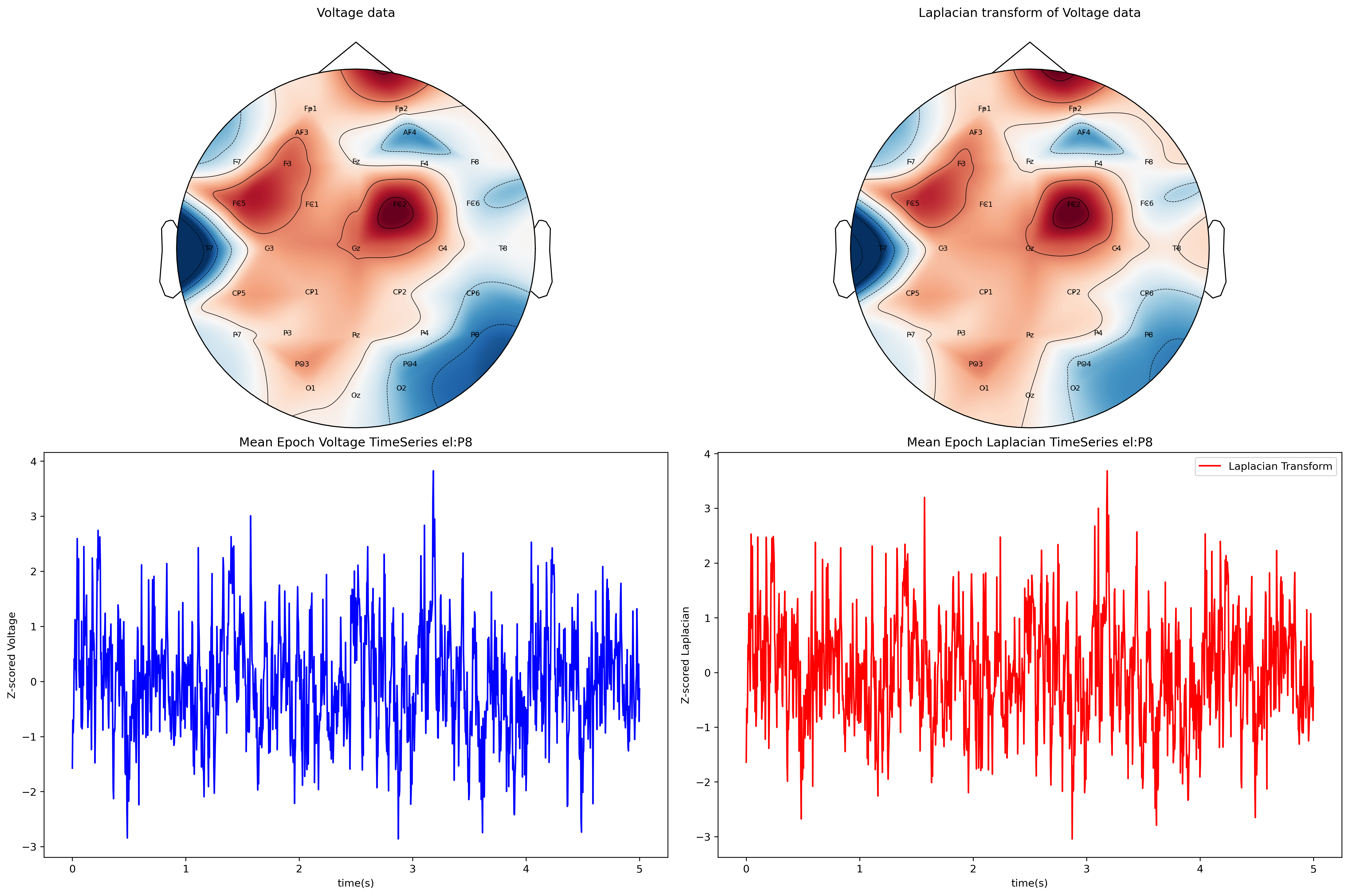}
	\captionsetup{justification=justified, labelfont=bf}
	\caption*{Supplementary Figure S3. Topographical voltage maps (top, plotted time point at 0.5s) and z-scored voltage time series (bottom) from a representative electrode (P8) before (left) and after Laplacian (right) spatial filtering of EEG data according to the method of Perrin (Perrin et al., 1989) to remove any volume conduction artifacts before connectivity analysis.}
\end{figure}

\begin{table}[!t]
	\centering
	\captionsetup{justification=justified, labelfont=bf}
	\renewcommand{\arraystretch}{1.3}
	
	\resizebox{\textwidth}{!}{%
		\begin{tabular}{|>{\centering\arraybackslash}m{3.5cm}|
				>{\centering\arraybackslash}m{4cm}|
				>{\centering\arraybackslash}m{4cm}|
				>{\centering\arraybackslash}m{4cm}|}
			\hline
			
			& \multicolumn{3}{c|}{Cliff’s delta or rank-biserial corr, CI [low, high], p-val FDR-corrected} \\
			\hline
			
			Laplacian Corrected Connectivity Metric & CN vs PDoff & CN vs PDon & PDoff vs PDon \\
			\hline
			
			Delta PLV & -0.542 [-1.82, -0.21], 0.0255 & -0.292 [-1.06, 0.26], 0.2081 & 0.6 [-0.47, 1.26], 0.2675 \\
			Theta PLV & -0.658 [-2.55, -0.57], 0.0090 & -0.642 [-2.15, -0.6], 0.0115 & -0.2 [-0.23, 0.51], 0.8904 \\
			Alpha PLV & -0.217 [-1.07, 0.63], 0.3041 & -0.175 [-1.01, 0.69], 0.4065 & 0.067 [-0.15, 0.35], 0.7984 \\
			Beta PLV & -0.3 [-1.37, 0.3], 0.1934 & -0.417 [-1.81, 0.12], 0.1202 & -0.333 [-0.85, -0.06], 0.2675 \\
			Gamma PLV & -0.375 [-1.48, 0.22], 0.1255 & -0.325 [-1.6, 0.43], 0.2053 & 0.067 [-0.41, 0.79], 0.7984 \\
			
			Delta PLI & -0.333 [-1.31, 0.28], 0.2845 & -0.242 [-0.95, 0.03], 0.7947 & 0.067 [-0.88, 0.61], 0.9780 \\
			Theta PLI & -0.092 [-0.9, 0.44], 0.6637 & 0.05 [-0.7, 0.77], 0.8433 & -0.2 [-0.26, 0.65], 0.9780 \\
			Alpha PLI & 0.183 [-0.46, 1.17], 0.5657 & 0.2 [-0.24, 1.29], 0.7947 & 0.2 [-0.03, 0.46], 0.4220 \\
			Beta PLI & 0.375 [0.01, 1.34], 0.2845 & -0.042 [-0.75, 0.68], 0.8433 & -0.733 [-1.46, -0.35], 0.0215 \\
			Gamma PLI & -0.158 [-1.14, 0.67], 0.5657 & -0.15 [-1.04, 0.63], 0.7947 & -0.067 [-0.58, 0.7], 0.9780 \\
			
			Delta wPLI & -0.408 [-1.39, 0.18], 0.1580 & -0.25 [-0.95, 0.06], 0.8125 & 0.2 [-0.85, 0.78], 0.9780 \\
			Theta wPLI & -0.142 [-0.98, 0.31], 0.5271 & -0.05 [-0.74, 0.65], 0.8125 & -0.067 [-0.23, 0.66], 0.9780 \\
			Alpha wPLI & 0.183 [-0.47, 1.19], 0.5271 & 0.2 [-0.28, 1.29], 0.8125 & 0.2 [-0.05, 0.43], 0.6310 \\
			Beta wPLI & 0.392 [0.01, 1.33], 0.1580 & -0.083 [-0.79, 0.63], 0.8125 & -0.6 [-1.59, -0.39], 0.0215 \\
			Gamma wPLI & -0.133 [-1.19, 0.65], 0.5271 & -0.142 [-1.08, 0.64], 0.8125 & -0.067 [-0.46, 0.73], 0.9780 \\
			
			\hline
		\end{tabular}%
	} 
	
	\caption*{Supplementary Table 1. Comparative summary of group-level comparisons using Phase-based connectivity metrics after Laplacian transform of EEG data across diagnostic contrasts. Values represent Cliff’s delta or matched-pairs rank-biserial correlation (unpaired or paired respectively) with 95\% confidence intervals and Wilcoxon (rank sum or signed rank sum) FDR-corrected p-values (Benjamini--Hochberg method). PLV, Phase Locking value; PLI, Phase Lag Index; wPLI, weighted Phase Lag Index; CN, Control; PDoff, Parkinson’s patient off medication; PDon, Parkinson’s patient on medication.}
\end{table}

\end{document}